\documentclass[aps, usenatbib,times]{mn2e}

\usepackage{amsmath, amssymb}
\usepackage{graphicx}

\usepackage{color}
\usepackage{graphicx}

\usepackage{dcolumn}
\usepackage{bm}

\newcommand{\dd}{{\rm d}}

\newcommand{\mK}{{\cal K}}

\newcommand{\mN}{{\cal N}}

\newcommand{\mS}{{\cal S}}
\newcommand{\mR}{{\cal R}}

\newcommand{\mG}{{\cal G}}

\newcommand{\beq}{\begin{equation}}
\newcommand{\eeq}{\end{equation}}
\newcommand{\beqa}{\begin{eqnarray}}
\newcommand{\eeqa}{\end{eqnarray}}

\def\la{\mathrel{\mathpalette\fun <}}
\def\ga{\mathrel{\mathpalette\fun >}}
\def\fun#1#2{\lower3.6pt\vbox{\baselineskip0pt\lineskip.9pt
        \ialign{$\mathsurround=0pt#1\hfill##\hfil$\crcr#2\crcr\sim\crcr}}}

\newcommand{\peff}{p_{\rm eff.}}

\newcommand{\tkappa}{\tilde{\kappa}}
\newcommand{\tw}{\tilde{w}}

\newcommand{\simlt}{\lower.5ex\hbox{$\; \buildrel < \over \sim \;$}}
\newcommand{\fK}{f_\mathrm{K}}
\newcommand{\gK}{g_\mathrm{K}}

\begin{document}


\title[Cosmic shear full nulling: sorting out dynamics, geometry and systematics]{Cosmic shear full nulling: sorting out dynamics, geometry and systematics}
\author[Francis Bernardeau, Takahiro Nishimichi, Atsushi Taruya ]{Francis Bernardeau$^{1,2}$, Takahiro Nishimichi$^{3}$, Atsushi Taruya$^{4,2,5}$ \\
$^{1}$ Institut de Physique Th\'eorique, CEA, IPhT and
CNRS, URA 2306, F-91191 Gif-sur-Yvette, France\\
$^{2}$
Research Center for the Early Universe, School of Science, 
The University of Tokyo, Bunkyo-ku, Tokyo 113-0033, Japan\\
$^{3}$
Institut d'Astrophysique
de Paris \& UPMC (UMR 7095), 98, bis boulevard Arago , 75 014, Paris,
France.\\
$^{4}$ Institute for the Physics and Mathematics of the Universe,
University of Tokyo, Kashiwa, Chiba 277-8568, Japan\\
$^{5}$Yukawa Institute for Theoretical Physics, Kyoto University, Kyoto 606-8502, Japan
}
\maketitle


\begin{abstract}
An explicit full nulling scheme for cosmic shear observations is presented. It makes possible the construction of shear maps from 
extended source distributions for which the \textsl{lens} distance distribution is restricted to a definite interval. 
Such a  construction allows to build totally independent shear maps, at all scales, that can be taken advantage 
of to constrain background cosmological parameters and systematics
using the full statistical power of cosmic shear observations. 
Another advantage of such construction is that, as the lens redshift distribution 
can be made arbitrarily narrow, scale mixing due to projection effects can be 
limited allowing controlled predictions on the  large scale shear power spectrum from perturbation theory calculations.\\
\end{abstract}

\section{Introduction}

After first detection of cosmic shear effects by \cite{2000Natur.405..143W,2000A&A...358...30V,2000MNRAS.318..625B} and the results obtained 
in more advance surveys (such as the CFHTLS survey, \cite{2008A&A...479....9F, 2012MNRAS.427..146H}) the science domain is about to enter an era of precision of large-scale measurements with a new generation of surveys either from ground-based facilities (e.g. DES,   Pan-STARRS, 
LSST\footnote{\texttt{https://www.darkenergysurvey.org},\,
\texttt{http://pan-starrs.ifa.hawaii.edu},\,\\
\texttt{http://www.lsst.org}}) 
or space-based observatories such as EUCLID\footnote{see~\cite{2011arXiv1110.3193L}.}.

Concurrently, a lot of efforts have been devoted to the development of analytical methods applied to the growth
of structure and in particular to the computation of power spectra beyond linear order. These methods try to improve upon standard perturbation theory
calculations and  aim at proposing first principle calculations of power spectra that are valid at significantly smaller scale
than standard linear theory. The first significant progress in this line of calculations is the RPT proposition \citep{2006PhRvD..73f3519C}
followed by the closure theory \citep{2008ApJ...674..617T} and the time flow equations approach proposed in \cite{2008JCAP...10..036P}. Latest
propositions, namely MPTbreeze \citep{2012MNRAS.427.2537C} and RegPT \citep{2012PhRvD..86j3528T} incorporate 2-loop order
calculations and are accompanied by publicly released codes. 
Provided  calculations are confined in their validity region, predictions from such codes can be extremely accurate, at percent level. It is then natural to try to apply these predictions to cosmic shear observations.  When applied to projected convergence maps however, the results are rather disappointing as projections effects tend to mix large and small scale. It then inevitably spoils the quality of the theoretical predictions.

We have identified however a way to circumvent this problem and it is based on a nulling approach, that is a method 
to reorganize the multi-source plane observations of cosmic shear in such a way that the redshift distribution of the sources can be manipulated at will. Nulling has been introduced in previous studies in \cite{2008A&A...488..829J} as a technique to circumvent intrinsic alignment effects by making the contributions of lenses null at a given redshift. So here we adopt a slightly different point of view. The point is not so much to find ways to circumvent such effects but to propose a transformation of the data that makes possible
to sort out the information content in the weak lensing observables.
This will be possible if the lens distribution can be confined
to a definite distance interval to avoid scale mixing. We will see that, solving this problem leads to a reorganization of the 
data in such a way that  most of the the cross spectra identically vanish for a given choice of a redshift-distance relation.
It allows us to apply perturbation calculations to analyze the data on large angular separations.
The nulling property of the transformed maps opens the path to pure geometrical tests that can be done 
without any knowledge of small scale physics.

Note that the aim of this study is similar to that of the 3D lensing technique 
(e.g., \citealt{2003MNRAS.343.1327H,2006MNRAS.373..105H,2011MNRAS.413.2923K}) 
in the sense that we are trying to extract the density fluctuations in three-dimensional wavenumber $k$ rather than 
the projected angular scales.
Also, several papers in the literature propose methods to avoid uncertainties on small scales 
\citep{2005PhRvD..72d3002H,2011MNRAS.416.1717K}.
This paper presents a simple method along these directions with a weighting scheme on galaxies 
according to their photometric redshift.

The plan of the paper is the following. In Section \ref{Sec:Discrete} we present the nulling solution for a set of discrete source planes as available in numerical simulations and exploit perturbation theory calculations to predict shear map
spectra and cross-spectra in this context. In Section \ref{Sec:Continuous} 
we present an alternative tomographic basis that exhibit nulling properties for continuous source
distributions. In Section \ref{sec:implication}  we explore the robustness of the nulling procedure when one introduces 
realistic statistical errors in the determination of the photometric redshifts and  when one varies the cosmological parameters.
We summarize our findings in the last section.

\section{The case of discrete source planes}
\label{Sec:Discrete}

The construction of full nulling selection function is particularly simple in case of discrete source planes. Let us then assume 
we have a discrete number of source planes at redshift $z_{i}$ at our disposal.  In general the local convergence $\kappa$ is given by
a line-of-sight integration given by (see for instance \cite{1999ARA&A..37..127M})
\begin{eqnarray}
\kappa&=&\frac{3\Omega_{0}H_0^2}{2c^2}\sum_{i} p_{i}\,\int_{0}^{\chi_{i}}\dd\chi\ \frac{\fK(\chi_{i}-\chi)\fK(\chi)}{\fK(\chi_{i})}
\frac{\delta(\chi)}{a(\chi)},
\label{kappaexpression}
\end{eqnarray}
where $\chi$ is the radial distance, $\chi_{i}$ is the radial distance to the redshift $z_{i}$, $K$ is the (constant) space curvature, $\delta(\chi)$ 
is the (total matter) density contrast along the line of sight,  $a(\chi)$ is the expansion factor and $p_{i}$ are 
dimensionless weight coefficients whose values
will be chosen in order to achieve the desired properties. 
In the above, we define the comoving angular diameter distance:
\begin{eqnarray}
\fK(\chi) \equiv\left\{
\begin{array}{cc}
\displaystyle\frac{\sin(\sqrt{K}\chi)}{\sqrt{K}} & \mathrm{for}\,\,K>0,\\
\displaystyle\chi & \mathrm{for}\,\,K=0,\\
\displaystyle\frac{\sinh(\sqrt{-K}\chi)}{\sqrt{-K}} & \mathrm{for}\,\,K<0.
\end{array}
\right.
\label{fKdef}
\end{eqnarray}
The expression (\ref{kappaexpression}) can be rewritten in the following form,
\begin{equation}
\kappa=\frac{3\Omega_{0}H_0^2}{2c^2}\int_{0}^{\chi_{\infty}}\dd\chi\ \frac{\delta(\chi)}{a(\chi)}\,w(\chi),
\end{equation}
with
\begin{equation}
w(\chi)=\sum_{i,\,\chi_{i}>\chi} p_{i}\,\frac{\fK(\chi_{i}-\chi)\fK(\chi)}{\fK(\chi_{i})},
\label{wchidef}
\end{equation}
where $\chi_{\infty}$ is the largest radial distance available and where the sum runs for source planes that are behind the lenses.
The function $w(\chi)$ here 
encodes the distance dependent weight with which lenses along the line of sight are contributing to the projected convergence.

The problem is now to choose a set of weights $p_{i}$ in order to build shear maps with a predefined weight form, $w(\chi)$, and
in case of discrete sources, in such a way that the lens distribution is confined in a finite range of distances. 
The mathematical solution for a set of discrete source planes turns out to be  non-ambiguous and well defined. 

\subsection{The 3 source plane solution}

To start with let us assume that we have 3 source planes at our disposal at given distances $\chi_{i}$, $i=1,3$. 
The expression of $w(\chi)$ can be fruitfully replaced by,
\begin{eqnarray}
w(\chi)&=&\fK^2(\chi)\left[\frac{1}{\gK(\chi)}\sum_{i, \chi_{i}>\chi}p_{i}-\sum_{i, \chi_{i}>\chi}\frac{p_{i}}{\gK(\chi_{i})}\right],
\label{w3}
\end{eqnarray}
where we have introduced
\begin{eqnarray}
\gK(\chi) \equiv\left\{
\begin{array}{cc}
\displaystyle\frac{\tan(\sqrt{K}\chi)}{\sqrt{K}} & \mathrm{for}\,\,K>0,\\
\displaystyle\chi & \mathrm{for}\,\,K=0,\\
\displaystyle\frac{\tanh(\sqrt{-K}\chi)}{\sqrt{-K}} & \mathrm{for}\,\,K<0.
\end{array}
\right.
\end{eqnarray}
Note that this result follows from the trigonometric identity of the sine and hyperbolic sine function 
in equation~(\ref{fKdef}).

The key remark underlying our paper is that if the weight associated with each source plane satisfies the 2 constraints, 
\begin{equation}
\sum_{i=1}^{3}p_{i}=0,\ \ \ 
\sum_{i=1}^{3}\frac{p_{i}}{\gK(\chi_{i})}=0,
\end{equation}
then
whenever $\chi<\chi_{1}$ we have $w(\chi)=0$ implying that the lenses all lie between $\chi_{1}$ and $\chi_{3}$. The previous conditions can be explicitly solved and one gets (to an arbitrary normalization),
\begin{eqnarray}
p_2 / p_1 =c(2,3,1)/c(1,2,3),\quad p_3 / p_1 =c(3,1,2)/c(1,2,3),
\label{solution3}
\end{eqnarray}
where
\begin{eqnarray}
c(i,j,k)=\gK(\chi_{i})\Bigl[\gK(\chi_{j})-\gK(\chi_{k})\Bigr].
\label{cijk}
\end{eqnarray}
Plugging this solution into equation~(\ref{w3}), we have the weighted lens distribution:
\begin{eqnarray}
w(\chi) = \left\{
\begin{array}{cc}
\displaystyle p_1\fK^2(\chi)\left[\frac{1}{\gK(\chi_1)}-\frac{1}{\gK(\chi)}\right] & \mathrm{for}\,\,\chi_1\leq \chi < \chi_2,\\
\displaystyle p_3\fK^2(\chi)\left[\frac{1}{\gK(\chi)}-\frac{1}{\gK(\chi_3)}\right] & \mathrm{for}\,\,\chi_2\leq \chi < \chi_3,\\
\displaystyle 0 & \mathrm{otherwise}.
\end{array}
\right.
\end{eqnarray}

The solution for the nulling condition is no longer unique when we have more than three source planes.
However, as discussed in the next section, we can still use equation~(\ref{solution3}) with three different indices 
chosen arbitrarily from the available source planes even in that case to construct nulling profiles.
The general solution can be obtained by taking the linear combinations of the three-plane solution~(\ref{solution3})
for different sets of planes.

\subsection{Resulting correlation structure for a set of discrete planes}
\label{lens4}
\begin{figure}
   \centering
 \includegraphics[width=7cm]{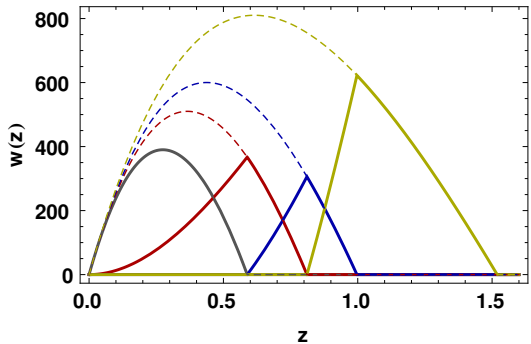} 
   \caption{Shape of the lens distribution $\tw_{a}(z)$ constructed from four lens planes with the weights of 
   Eq. (\ref{weightmatrix}) corresponding to the simulation of \citet{2009ApJ...701..945S} with $a=1$ to $4$
   from left to right (see section~\ref{lens4} for detail).
   The lens redshift distributions $\tw_{a}(z)$ and $\tw_{b}(z)$ do not overlap when $\vert a-b\vert \geq 2$.
   Note that the y-axis is shown in unit of $h^{-1}\mathrm{Mpc}$.}
   \label{pofznum}
\end{figure} 

If we have a larger set of discrete planes
we can define an ordered set of source distributions for which the resulting cosmic shear maps are correlated only to their nearest ones. 
So let us consider a set of $n$ discrete source planes $\kappa_{i}$ located at $\chi_i$ where $i=1, 2, \dots, n$
and $\chi_i<\chi_{i+1}$. 
One can define the $n$ maps $\tkappa_{a}$ by taking 
linear combinations of the original maps $\kappa_{i}$, 
\begin{equation}
\tkappa_{a}=\sum_{i}p_{a}^{i}\ \kappa_{i}.
\end{equation}

We can pick three neighboring lens planes and apply
the three-lens solution~(\ref{solution3}) to have a nulling profile.
We can construct $(n-2)$ new maps with nulling implemented by doing this to every set of three neighboring planes.
We label them as $\tkappa_{a}$ with $a\geq3$, and these maps are nonzero between $\chi_{a-2} < \chi < \chi_{a}$.
We add two more maps, $\tkappa_1=\kappa_1$ and $\tkappa_2=\kappa_2-\kappa_1$, to have a complete set of
planes without loosing any information in the original $n$ planes (i.e., the matrix $p_{a}^{i}$ is invertible).
Note that these two planes have nonzero lensing response down to $z=0$ as we do not apply the nulling condition to them.
With this labeling convention, we have $n$ new maps $\tkappa_a$, whose covering redshift ranges are 
in ascending order of $a$. 

To summarize, the non zero coefficients are given by
\footnote{We impose here a choice of normalization so that the diagonal of the matrix contains only 1.},
\begin{eqnarray}
p^{1}_{1}=1,\qquad p^{1}_{2}=-1,\qquad p^{2}_{2}=1,
\end{eqnarray}
for the first two maps ($a=1$ and $2$), and the remaining maps $\tkappa_{a\geq3}$ are given by
\begin{eqnarray}
p^{a}_{a}&=&1,\\
p^{a-1}_{a}&=&c(a-1,a-2,a)/c(a-2,a,a-1),\\
p^{a-2}_{a}&=&c(a-2,a-1,a)/c(a-2,a,a-1),
\end{eqnarray}
where $c(i,j,k)$ is defined in Eq.~(\ref{cijk}).
It is important to note that
this transformation of the $\kappa_{i}$ maps into $\tkappa_{a}$ maps is regular. As a consequence, it does not change their information content. What we have gained here, as we will illustrate in the following, is to partially sort out the information content
of the maps. It is done in two ways,
\begin{itemize}
\item starting with $a=3$ the maps are built out of a finite range in redshift;
\item the lens distributions for $\tkappa_{a}$ and $\tkappa_{a+2}$ do not overlap.
\end{itemize}

We can illustrate this construction with a simple example we will exploit in the following to compare our results with numerical 
simulations. Using the simulations provided by \cite{2009ApJ...701..945S}, we can exploit up to six source planes but will
restrict our analysis here to the first four at $z=0.589,\ 0.811,\ 0.999$ and $1.52$ (just in order to be realistic). For a flat universe 
with $\Omega_{0}=0.238$, the distances to the source planes are $\chi=0.520,\  0.680, 0.800$ and $1.080 $ in units of 
$c/H_{0}$. The resulting weight matrix reads,
\begin{eqnarray}
p_{a}^{i}=
\left(
\begin{array}{cccc}
 1 & 0 & 0 & 0 \\
 -1 & 1 & 0 & 0 \\
 0.4875 & -1.4875 & 1 & 0 \\
 0 & 1.46911 & -2.46911 & 1
\end{array}
\right)
\label{weightmatrix}
\end{eqnarray}
and the resulting lens weight function $\tw_{a}(z)$ are shown on Fig. \ref{pofznum} (solid lines).
We also show the original profiles in dashed line before implementing nulling. 
Two such resulting convergence maps 
with indices that differ by more than 2 are, to systematic error effects, totally independent.

\subsection{Predictions from Perturbation Theory calculations}
\label{Sec:PTresults}

\begin{figure}
   \centering
 \includegraphics[width=7cm]{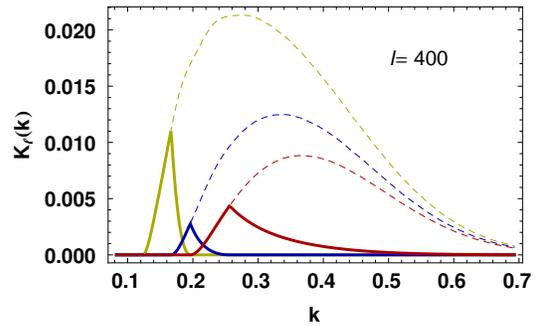} 
   \caption{The kernel $\mK_{\ell}(k)$ defined in Eq. (\ref{KernelDef}) showing the contributing values of $k$ for $\ell=400$ for the profiles 
   2 to 4 (same color coding as on Fig. \ref{pofznum}). 
   The solid lines are for the nulling profiles and the dashed lines are for the corresponding source planes without nulling. 
   Note that the y-axis is shown in unit of $h\,\mathrm{Mpc}^{-1}$.}
   \label{kernelklprof4}
\end{figure} 

We have reached here the original goal of this construction as the mapping between $\ell$ and $k$
values is now much better behaved than in standard tomographic approaches. 
This is illustrated on Fig.~\ref{kernelklprof4} (the precise definition of the kernels is given below) that shows that the contribution
to $C_{\ell}$ for a given $\ell$ is now restricted to a finite range of $k$. 
We are now in position to fruitfully apply perturbation 
theory results to the projected convergence maps that are constructed through this procedure.

In the following we will compare results of numerical simulations with prediction of the RegPT scheme described in
\cite{2012PhRvD..86j3528T} 
at 1-loop and 2-loop order.  We will also compare the results obtained when nulling is applied or not.
The RegPT scheme is based on some resummation properties of the propagators and it is beyond the scope 
of this paper to give a detailed presentation of it. We refer the reader to \cite{2012PhRvD..86j3528T}  for a detailed 
presentation of this scheme and how it differs from other possible approaches.
We simply recall that RegPT results can be reconstructed from standard PT diagrams. Each of these diagrams has a simple 
time dependence and the global time dependence of the power spectra can then easily be reconstructed from the 
results of the execution of the code \textsc{RegPT} (see again \cite{2012PhRvD..86j3528T}  for detail).

The $C_{\ell}^{ab}$ cross-power spectra are then computed from the relation,
\begin{equation}
C_{\ell}^{ab}=\int\dd \chi\ \mK_{\ell}\left(\frac{\ell}{\fK(\chi)}\right),
\label{eq:power}
\end{equation}
with
\begin{equation} 
\mK_{\ell}(k)=\frac{9\Omega_{0}^{2}H_0^4}{4c^4}
\,P\left(k,\eta(\chi)\right)\frac{w_{a}(\chi)w_{b}(\chi)}{a(\chi)^{2}
\fK(\chi)^2
},
\label{KernelDef}
\end{equation}
where $w_{a}(\chi)$ and $w_{b}(\chi)$ are the lens distribution functions for the shear maps $(a)$ and $(b)$
and $P(k,\eta)$ is here the linear, 1-loop or 2-loop order RegPT power spectrum as a function of time.
Figure \ref{kernelklprof4} shows example of kernels in $k$ that contribute to values of $C_{\ell}$ for a given 
value of $\ell$ and for profiles 2, 3 and 4. 

\begin{figure}
   \centering
 \includegraphics[width=7cm]{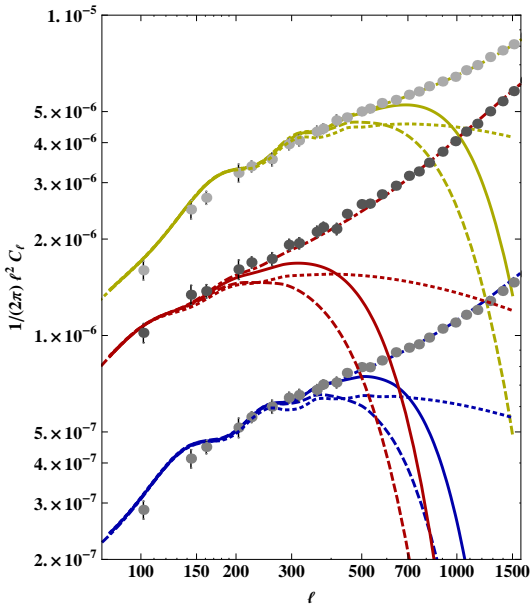} 
   \caption{The resulting power spectra for the second, third and fourth bin (same color coding as on Fig. \ref{pofznum}) when
   nulling is implemented. We plot the predictions of linear theory, RegPT at the 1- and 2-loop order and the revised halofit 
   respectively by the dotted, dashed, solid and dot-dashed line,
   while the measurement from the simulation by \citet{2009ApJ...701..945S} is shown by symbols.}
   \label{SatoCl_Nulling}
\end{figure}

\begin{figure}
   \centering
 \includegraphics[width=7cm]{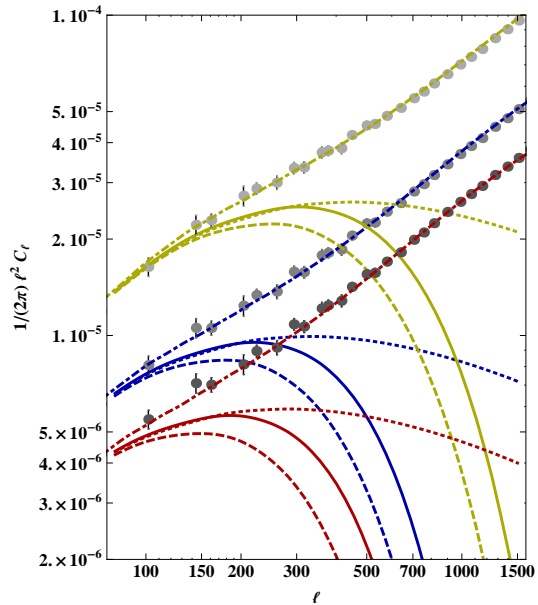} 
   \caption{The resulting power spectra for the second, third and fourth source planes (same color coding as on Fig. \ref{pofznum}) when
   nulling is not implemented. See text for details.}
   \label{SatoCl_NoNulling}
\end{figure} 

\begin{figure}
   \centering
 \includegraphics[width=7cm]{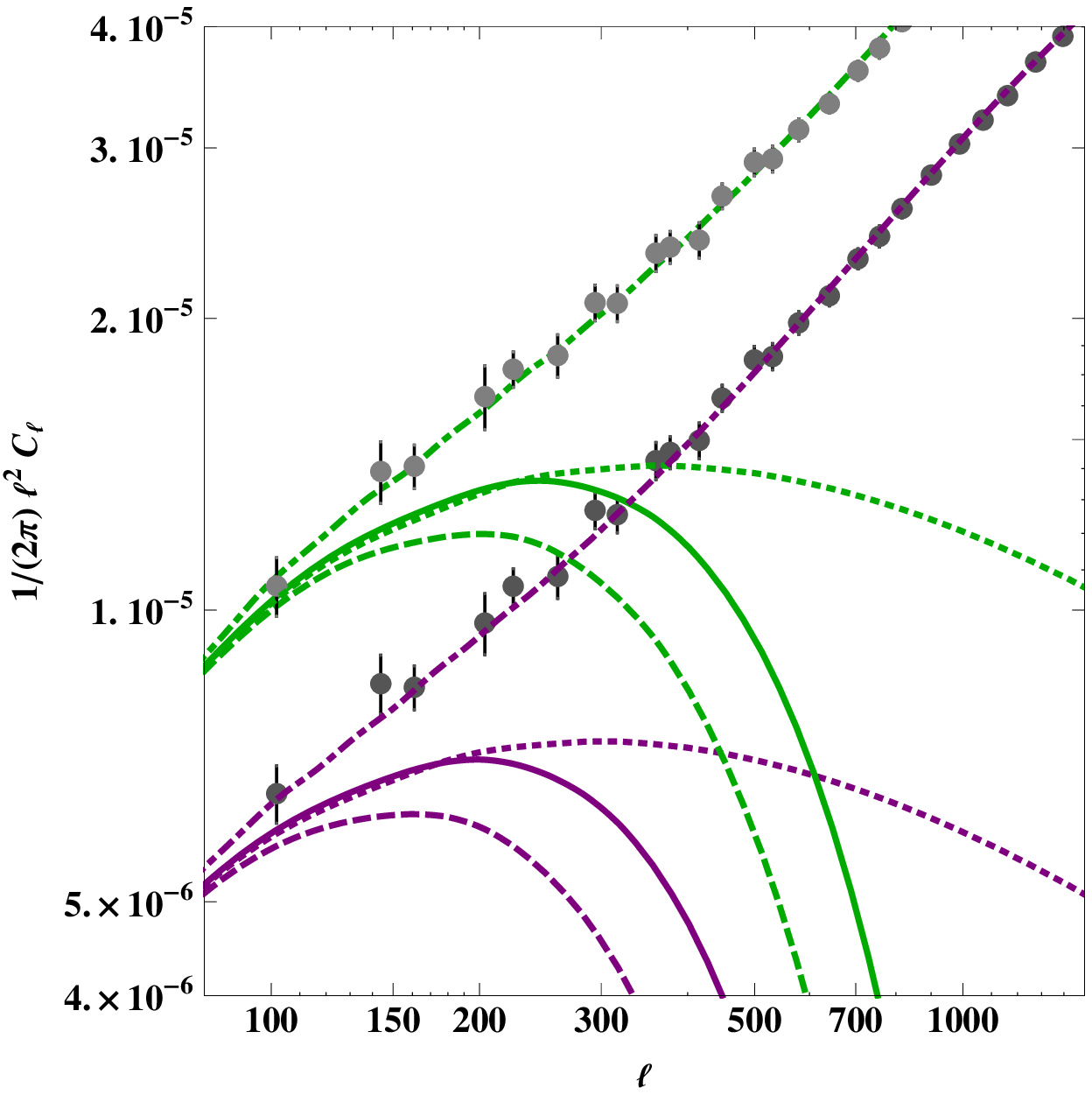} 
 \includegraphics[width=7cm]{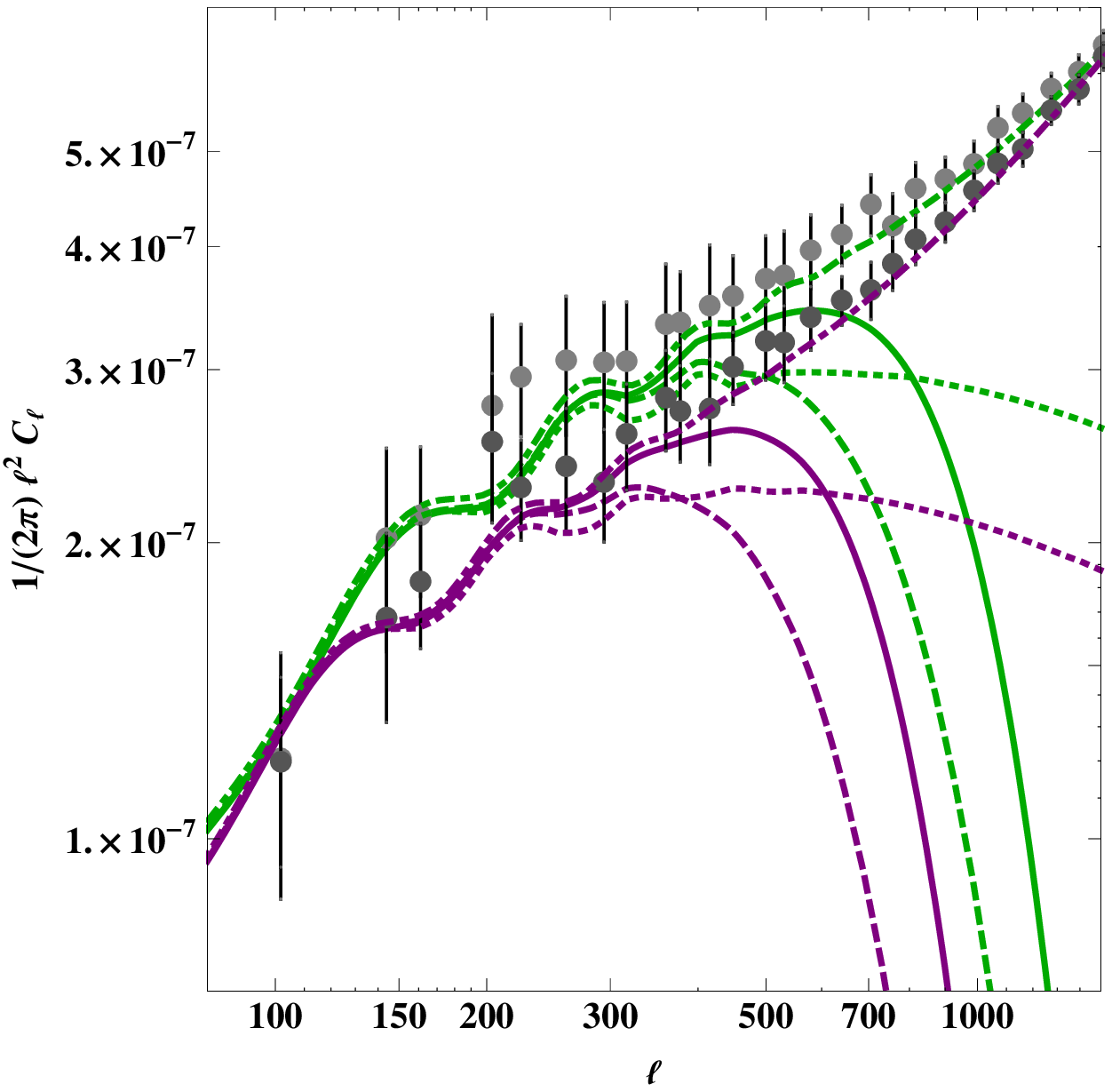} 
   \caption{The resulting cross power spectra between two subsequent bins, the second and third and the third and fourth. Top panel is 
   when no nulling is implemented and bottom panel with nulling.}
   \label{SatoXl_Nulling}
\end{figure}

We can then compute the resulting cross-spectra matrix for a set of 4 redshifts. As mentioned before, the cross-spectra 
matrix is band diagonal. To illustrate the performance of the computation we present the auto-correlation function
for the third and fourth bin (corresponding to realistic redshift ranges) on Fig. \ref{SatoCl_Nulling}. The dotted line is 
the linear theory, the dashed line is the 1-loop order RegPT result, the solid line is the 2-loop order
RegPT result and the dot-dashed line is fitting formula for the nonlinear power spectrum 
(halofit: \citealt{2003MNRAS.341.1311S}) with revised parameters calibrated in \citet{2012ApJ...761..152T} 
(the revised halo fit, hereafter).
Plotted in symbols are the measurements from the ray-tracing simulation by \cite{2009ApJ...701..945S}
with error bars showing the one-$\sigma$ statistical uncertainty estimated from the scatter among the 1000 independent
random realizations.
We here take full advantage of
the nulling prescription as it allows to extend the validity regime of perturbation theory calculations to values of
$\ell$ of about 1000. This is to be compared to standard linear regime prediction which are valid to $\ell$ of about
100 as shown on Fig. \ref{SatoCl_NoNulling} in the absence of nulling and for which PT predictions appear very 
poor because of scale mixing.

Finally in Fig. \ref{SatoXl_Nulling} we present similarly the cross-spectra between those two nearby bins 
(bins that are 2 indices apart exhibit of course no correlations at all). As for the auto-correlation spectra, 
the contribution for such cross-spectra is restricted in redshift.  In all cases predictions are compared
to the results of the numerical experiment of \citet{2009ApJ...701..945S}. 
Note however large-scale discrepancy between the predictions and the measurements due to finite area effects
(see Appendix~\ref{app:window} for detail).

Note that the revised halofit gives a good prediction over the plotted scale.
This does not come as a surprise since the fitting formula is indeed calibrated by $N$-body simulations 
conducted with the same numerical codes with similar simulation parameters to the one shown here.
Of course, further calibrations with refined simulations and eventual inclusion of possible impact
of baryonic physics into the fitting formula are natural steps forward. Our strategy presented here is heading
towards another direction: we are sorting out the cosmological information contents in nonlinear 
(complex and uncertain) regime from linear (clean and robust) regime, 
where the latter is accessible with perturbative techniques without free parameters or calibrations from N-body results.

\section{Nulling with realistic data sets}
\label{Sec:Continuous}

When the sources are not confined in discrete source planes,  the function $w(\chi)$, defined in eq. (\ref{wchidef}), is to  be computed
from a continuous source distribution,
\begin{equation}
w(\chi)=\int_{\chi}^{\chi_{\infty}}\dd\chi_{s}p(\chi_{s})n(\chi_{s})\frac{\fK(\chi_{s}-\chi)\fK(\chi)}{\fK(\chi_{s})}
\label{eq:wkernel}
\end{equation}
where $\chi_{\infty}$ is the largest (finite) accessible distance
to the observer and  $n(\chi_{s})$ is the given distance source distribution (which can be transformed into a redshift source distribution) provided
by the characteristics of the survey.  Following the standard ideas of the tomographic analysis \citep{1999ApJ...522L..21H}, the
point is to select sources
in redshift bins to gain information on the redshift evolution of clustering. In this case we can introduce the
function $p(\chi_{s})$ that can then be viewed as a free 
parameter that the observer is free to adjust to one's needs. The function $p(\chi_{s})$ will depend on the choices of boundaries
$\chi_{1}$ and $\chi_{2}$ within which we require the lens distance  to be bounded.

\subsection{The multi-plane solution and the continuous limit}

\begin{figure}
   \centering
 \includegraphics[width=7cm]{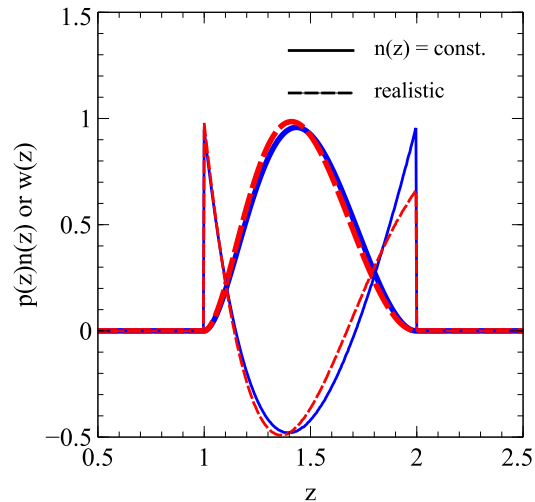}
   \caption{Example of source number density profiles (thin lines) and corresponding lens profiles (thick lines)
   constructed over a redshift interval of $1<z<2$. The solid (blue) lines correspond to the solution (\ref{eq:nchishape1}) 
   in case of a constant source number density, while
   the dashed (red) lines show the results for $n(z)$ in Eq.~(\ref{eq:nz}).
   Note that $p(z)n(z)$ ($w(z)$) has a dimension of 1/length (length), and their normalization can be taken arbitrary.
}
   \label{fig:prescription1}
\end{figure}

The constraints for $p(\chi)$, that we note $p(\chi_{s};\chi_{1},\chi_{2})$ in the following, one wishes to satisfy are then the following,
\begin{eqnarray}
\int_{\chi_{1}}^{\chi_{2}}\dd\chi_{s}p(\chi_{s};\chi_{1},\chi_{2})n(\chi_{s})&=&0,\label{CNulling1}\\
\int_{\chi_{1}}^{\chi_{2}}\dd\chi_{s}\frac{p(\chi_{s};\chi_{1},\chi_{2})n(\chi_{s})}{\gK(\chi_{s})}&=&0\label{CNulling2}
\end{eqnarray}
in order to meet the requested constraints.
The difference between the discrete case is that there is 
a whole set of continuous solutions to this system. 
Further constraints should then be imposed  in order to obtain a well defined solution and the natural constraint to put
is to maximize the signal to noise. In particular we do not want the resulting lens selection function to be too much 
oscillatory making the signal small and the noise too large.
The key is then to define a realistic prescription for the signal to noise. 
While the noise can be determined for a given survey setting (the galaxy redshift distribution, more specifically),
the signal can in principle freely be designed depending on the scale of interest and the statistical quantity one considers.
Thus, a fully valid prescription is non trivial in the sense that 
it should make intervene the nonlinear growth of perturbation which in turn is scale dependent 
(see Appendix~\ref{app:prescription} for some example prescriptions taking account of the nonlinear growth). 

A simple
prescription is to assume that the density contrast is simply a factor that grows like the expansion factor $a(\chi)$.
In this case, the convergence field scales as (see Eq.~\ref{kappaexpression})
\footnote{Note that $\chi_{2}$ could be set to an arbitrarily large value in the equations we are manipulating.}
\begin{eqnarray}
\mS_{1}&\propto&\int_{0}^{\chi_{\infty}}\dd\chi\,
w(\chi),
\nonumber\\
&=&\int_{\chi_1}^{\chi_2}\dd\chi_{s}p(\chi_{s})n(\chi_{s})\int_{0}^{\chi_{s}}\dd\chi \frac{\fK(\chi_{s}-\chi)\fK(\chi)}{\fK(\chi_{s})},
\label{eq:S1}
\end{eqnarray}
independently of the multipole.
We adopt $\mS_{1}$ as a simple estimate of the signal in the optimization.
Note that the $\chi$-integral can analytically be done and the result depends on the sign of the curvature.
As for the noise, we adopt the scaling 
\begin{eqnarray}
\mN^{2}\propto\int_{\chi_{1}}^{\chi_{2}}\dd\chi_{s}\,p^{2}(\chi_{s})n(\chi_{s}),
\label{eq:N}
\end{eqnarray}
which refers to the intrinsic shape noise contamination to the power spectrum.
With these expressions for the signal and the noise, one can find explicit forms for $p(\chi)$ that satisfies 
the constraints (\ref{CNulling1}) and (\ref{CNulling2}) and maximize the signal to noise ratio for a flat universe.
The ratio, which we denote by $\mR$, can be rewritten in a simple form:
\begin{eqnarray}
\mR = \frac{\mS_1}{\mN} \propto \frac{\left(p\cdot \chi^{2}\right)}{\left(p\cdot p\right)^{1/2}},
\end{eqnarray}
where we denote a scalar product of functions by
\begin{eqnarray}
\left(g_{1}\cdot g_{2}\right)\equiv \int_{\chi_{1}}^{\chi_{2}}n(\chi)\dd\chi\,g_{1}(\chi)g_{2}(\chi).
\end{eqnarray}
The constraints (\ref{CNulling1}, \ref{CNulling2}) now take the form,
\begin{eqnarray}
\left(p \cdot 1\right)=0,\ \ \left(p\cdot 1/\chi\right)=0.
\end{eqnarray}

Maximizing $\mR$ amounts then to find the function $p(\chi)$ in the subspace orthogonal to $m_{0}$ 
and $m_{-1}$ with the largest possible component along $m_{2}$, where $m_{i}=(p\cdot\chi^i)/(p\cdot p)^{1/2}$.
The solution is obtained as the result 
of a simple projection operator. More specifically the 
resulting source plane distribution takes the form (again, to an arbitrary normalization),
\begin{equation}
p(\chi;\chi_{1},\chi_{2})=\frac{1}{\chi_1\chi_2}\left[p_{2}(\chi)-\frac{\left(p_{2}\cdot 1/\chi\right)} {\left(1/\chi\cdot 1/\chi \right)}\ p_{-1}(\chi)\right],
\label{eq:nchishape1}
\end{equation}
where
\begin{equation}
p_{\alpha}(\chi)=\chi^{\alpha}-\frac{\left(\chi^{\alpha}\cdot 1\right)}{\left(1 \cdot 1\right)}.
\end{equation}
An explicit solution can be found if the available source distribution $n(\chi)$ is flat. It is then given by,
\begin{eqnarray}
p(\chi;\chi_{1},\chi_{2})&=&
\frac{\chi ^2}{\chi_1\chi_2}-\frac{1}{3} \left(\frac{\chi_1}{\chi_2}+\frac{\chi_2}{\chi_1}+1\right)\nonumber\\
&&-\frac{1}{6}\left(\frac{\chi_2-\chi_1}{\chi}+\log \left(\chi_2/\chi_1\right)\right) \nonumber\\  
&&\hspace{-2.cm}\times {\left[3\left(\frac{\chi_2}{\chi_1}-\frac{\chi_1}{\chi_2}\right)-2 \left(\frac{\chi_1}{\chi_2}+\frac{\chi_2}{\chi_1}+1\right) \log\left(\frac{\chi_2}{\chi_1}\right)\right]}\nonumber\\
&&\hspace{-2.cm}/\,\left[\left(\frac{\chi_1}{\chi_2}+\frac{\chi_2}{\chi_1}-2\right)-\left(\log\left(\frac{\chi_2}{\chi_1}\right)\right)^2\right].
   \label{eq:nchishape1_const}
\end{eqnarray}

We show in Fig. \ref{fig:prescription1}, the resulting form of the weighted source distribution 
$p(\chi)n(\chi)$ and the lens distribution $w(\chi)$ as a function of redshift (solid lines).
We also plot the solution~(\ref{eq:nchishape1}) as well as the corresponding lens distribution 
by dashed lines when a realistic redshift distribution of source galaxies is adopted 
(see Eq.~\ref{eq:nz} below).
The shape of the weighted source distribution is very similar in the two cases
and is regular enough to be constructible from actual source distribution. Note though
that it exhibits sharp features, discontinuities
at the both ends, $\chi_{s}=\chi_{1}$ and $\chi_{s}=\chi_{2}$.  
In the case with a realistic source distribution
the distribution at high redshift ($z\sim2$) is smaller than in the constant case, reflecting the fact that the (unweighted) source
number density is a decreasing function of $z$ over this redshift range. 
If one overweights the high-$z$ end, the resultant shape noise becomes relatively more important in the constructed map.
Our signal-to-noise maximization scheme works in this way and therefore controls the relative weight around the 
high- and low-redshift ends.

We can choose different signal-to-noise prescriptions to determine the shape of the weight function 
that satisfies Eqs.~(\ref{CNulling1}) and (\ref{CNulling2}). Although one cannot express the solution 
analytically in general, one still can solve them numerically with a reasonable choice of prescription. 
We explore some other prescriptions and summarize the results in Appendix~\ref{app:prescription}. 
Since it turns out that the resultant weight function is not very sensitive to
the prescription of the signal to noise, we simply adopt Eq.~(\ref{eq:nchishape1}) in what follows.

\subsection{Construction of a basis of source planes}

\begin{figure}
   \centering
 \includegraphics[width=7cm]{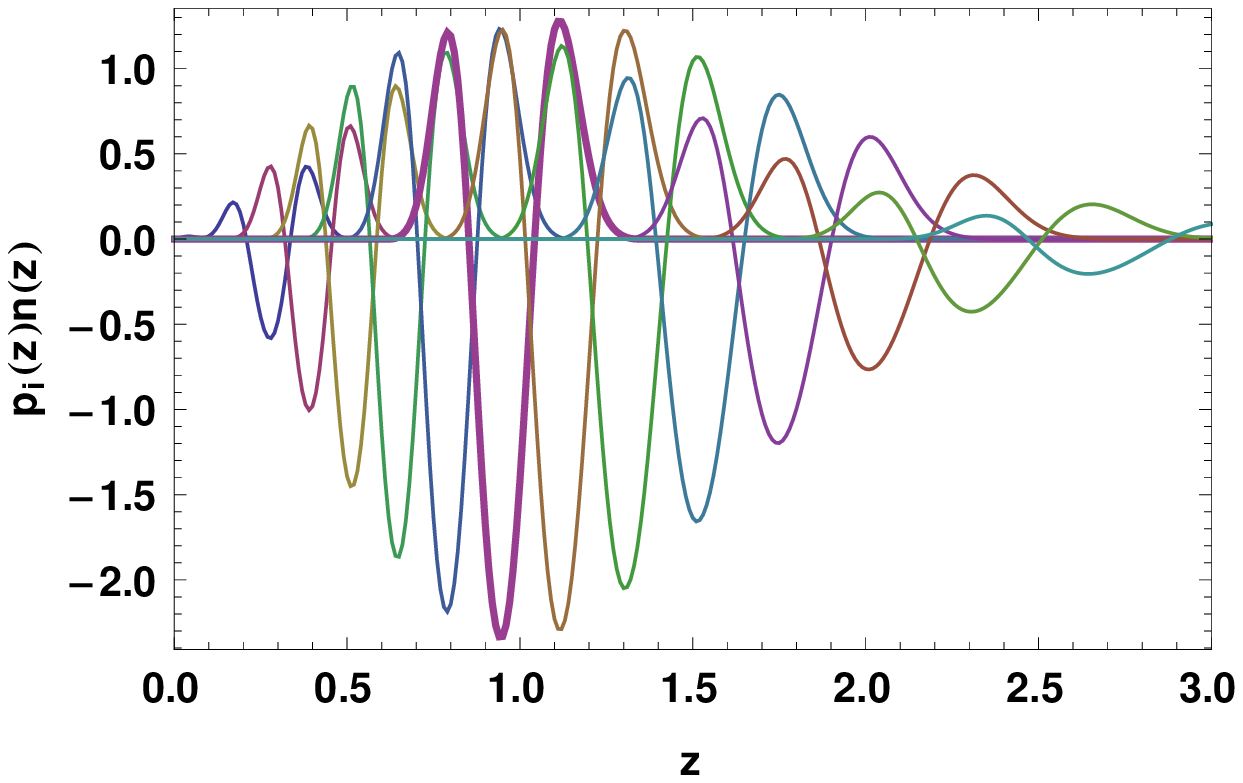} \hspace{1cm}
 \includegraphics[width=7cm]{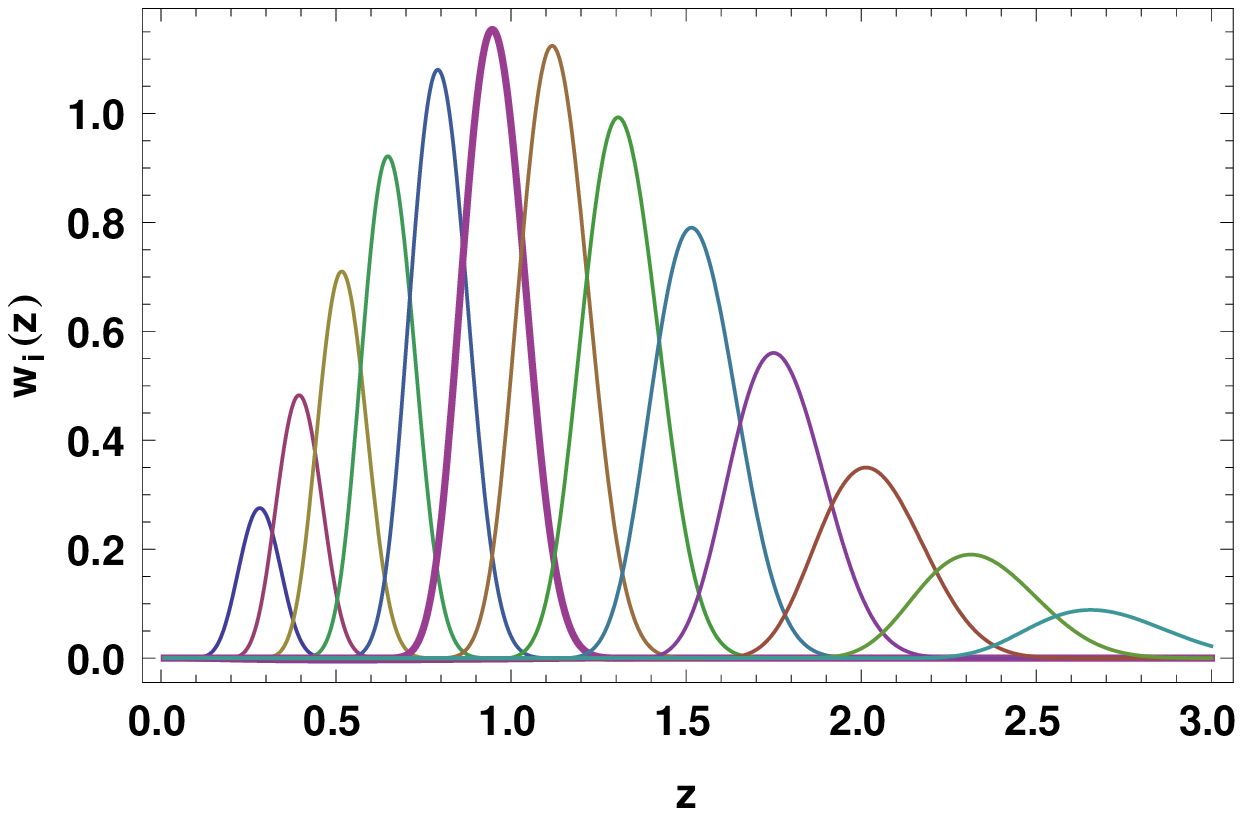} 
   \caption{The adopted profiles in redshifts for the sources (top panel) and the resulting profiles for the lens distribution 
   (bottom panel) for the fiducial cosmological model. The profiles for the sources have been obtained from the form 
   (\ref{eq:nchishape1}) after they are convolved
   with a kernel that mimic the photometric redshifts error distribution.
   Note that $p_i(z)n(z)$ ($w_i(z)$) has a dimension of 1/length (length), and their normalization can be taken arbitrary.
   \label{Profiles}}
\end{figure} 

The previous construction is still artificial in the sense that it assumes an infinite number of tracers. 
For realistic data sets,  one should take into account not only the continuous nature of the source 
distribution  but also the finite number of sources and the errors in the redshift determination. As a 
result it is clearly illusory to define arbitrarily narrow source distribution nor source distributions with 
too sharp features. It is  possible however to obtain smooth source distributions from superpositions
of $p(\chi_{s},\chi_{1},\chi_{2})$ taking advantage of the linearity of the constraints (\ref{CNulling1}-\ref{CNulling2}). 
As a consequence, one can convolve $p(\chi_{s},\chi_{1},\chi_{1}+\Delta\chi_{1})$ with any 
kernel function  $\mG(\chi_{1}\!-\!\chi'_{1})$  broader or of width comparable to the typical expected 
error distribution in the distance. We  can then build a set of profiles as
\begin{eqnarray}
\peff(\chi_{s};\chi_{1},\chi_{1}+\Delta\chi_{1})&=&\nonumber\\
&&\hspace{-2cm}\int\dd\chi'_{1}\,p(\chi,\chi'_{1},\chi'_{1}+\Delta\chi_{1})\ \mG(\chi_{1}\!-\!\chi'_{1}).
\label{neff}
\end{eqnarray}
with arbitrary values for $\chi_{1}$ and $\Delta\chi_{1}$ that determine respectively the overall distance 
to the sources and its width. By linearity, the resulting shape preserves the nulling property of the original distribution.
An example of such a profile is presented on Fig. \ref{Profiles}, top panel thick line, with the corresponding lens distribution
(bottom panel, thick line) where we use for $\mG(\chi_{1}\!-\!\chi'_{1})$ a kernel that 
corresponds to a $(1+z)3\%$ dispersion in the redshift determinations.

It is possible to vary $\chi_{1}$ to build a whole set of nulling functions that can form a basis on which to analyze the data.
We propose on Fig. \ref{Profiles} an explicit construction of such functions. Here we assume for the total source
distribution, 
\begin{equation}
n(z)\sim(z/z_{0})^{2}\exp[-(z/z_{0})^{1.5}]
\label{eq:nz}
\end{equation}
with $z_{0}=0.8$.
The functions are regularly spaced in radial distances, i.e. $\chi_{i}=(0.05+i)c/H_{0}$ for $i=1,13$ 
and $\Delta\chi_{1}=0.2\ c/H_{0}$. 
They are found to be smooth enough to be constructible from a realistic $z$ distribution. 
The nulling property for this choice of functions is clearly visible on the bottom panel as the lens distributions 
are seen to be restricted into definite intervals.

Clearly such functions can serve as a basis for the source profiles.  It can indeed be used to 
reconstruct any source distribution with the 
observed redshift resolution.  One can then replace standard tomographic binning 
by a finite set of such functions with no loss of information. We leave for further studies the description 
of an optimal choice of basis.

\section{Implications}
\label{sec:implication}
\subsection{Accuracy of nulling in realistic situations}
\label{subsec:accuracy}
There are two sources that in practice prevent us from a perfect nulling. Firstly, any dispersion of the photometric redshift
widens the lensing profile as we have already discussed. Secondly, nulling requires the background geometry of
the universe between the source galaxies and us to be known as an incorrect assumption in the cosmological model
leads to a failed nulling profile. In this subsection, we quantify the imperfectness of nulling from these two effects
employing two adjacent profiles which do not overlap when the nulling is perfect, and discuss the requirements
to achieve successful nulling properties.

We consider a redshift interval of $1<z<2$ and implement nulling to the source
galaxies in this (photometric) redshift range with various assumptions.
We consider the source distribution function given by Eq.~(\ref{eq:nz}), and
adopt Eq.~(\ref{eq:nchishape1}) to construct a smooth profile.
If nulling is implemented successfully, the resulting lensing profile 
should be consistent with zero at lower redshifts (i.e., $z<1$). We prepare another profile
to cover $0<z<1$ and check whether the nulled profile really does not respond to the structure between the observer 
and $z=1$ by taking the cross correlation of the two profiles.
We construct the second profile by giving a uniform weight over the source galaxies at $z<1$ for simplicity.

\begin{figure}
   \centering
 \includegraphics[width=8.5cm]{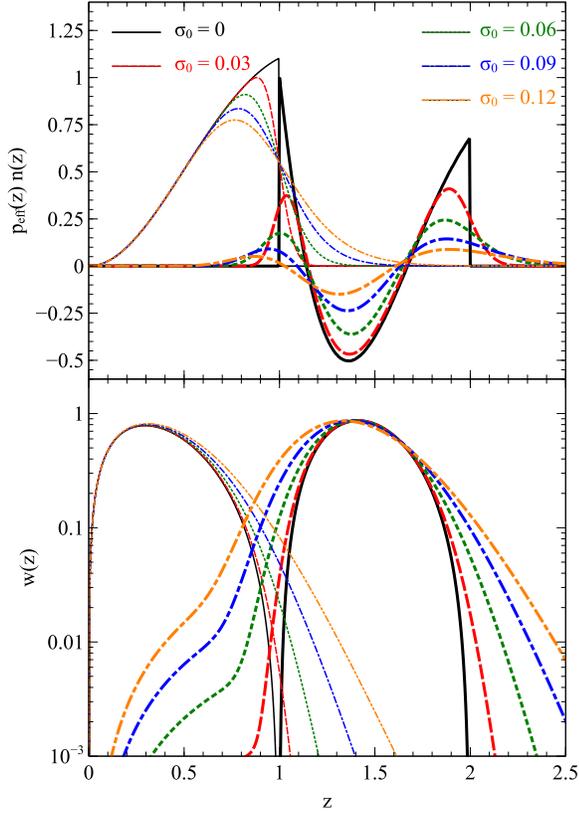}
   \caption{Effect of the dispersion of the photometric redshifts on the nulling. Two adjacent profiles (thin: $0<z<1$, 
   thick: $1<z<2$) are plotted for different values of the dispersion parameter, $\sigma_0$, as shown in the legend.
   We plot the weighted number density of source galaxies (top panel) and the resultant lensing profile (bottom).
   Note that $p_\mathrm{eft}(z)n(z)$ ($w(z)$) has a dimension of 1/length (length), 
   and their normalization can be taken arbitrary.
}
   \label{fig:kernel_photoz}
\end{figure}

The weighted source number density, $p_\mathrm{eff}(z)n(z)$, for the two profiles are shown in the top panel of 
Fig.~\ref{fig:kernel_photoz} when the dispersion of the photometric redshift is given by $\sigma_z(z) = \sigma_0(1+z)$ 
with various values of $\sigma_0$ assuming a Gaussian photometric redshift distribution; 
$\sigma_0 = 0$, $0.03$, $0.06$, $0.09$ and $0.12$ for solid, 
dashed, dotted dot-dashed dot-dot-dashed line, respectively. We plot the profile implementing nulling by thick lines
while the other profile covering $0<z<1$ is depicted by thin lines.
Note that the distribution of the photometric redshift can be much more complicated in reality with 
e.g., catastrophic errors or redshift dependent dispersion.
The bottom panel shows the lensing profile $w$ (see equation~\ref{eq:wkernel} for the definition) 
corresponding to the source distribution in the top panel in the same line type.
The two profiles approach zero at $z=1$ when we do not consider the dispersion in photo-$z$ 
(i.e., $\sigma_0=0$; solid).
For increasing the value of $\sigma_0$, the overlap between the two becomes significant. 

\begin{figure}
   \centering
 \includegraphics[width=7cm]{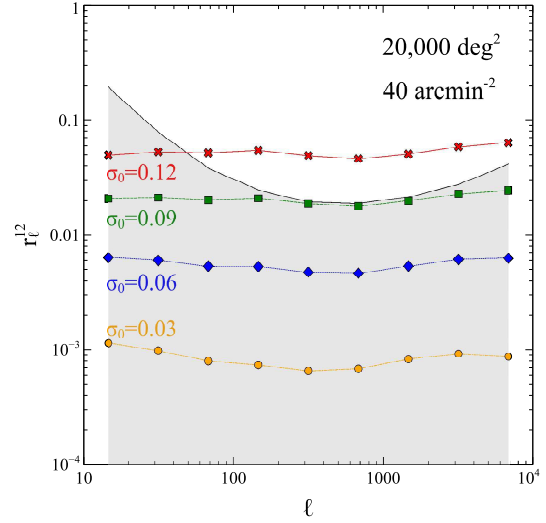}
   \caption{Cross correlation coefficient, $r_\ell^{12}$, for different values of the dispersion of the
   photometric redshifts as shown in the figure legend. The statistical error on this coefficient (i.e., the 
   cosmic variance and the
   shape noise, see equation~\ref{eq:error12}) is shown by the shaded region for
   a survey with $n_\mathrm{tot} = 40\,\mathrm{arcmin}^{-2}$ in a area of $20,000\,\mathrm{deg}^2$.
   Note that the error level explicitly depends on the bin size of $\ell$ (see equation~\ref{eq:error12} and \ref{eq:Nmode}),
   and we adopt a logarithmic binning with three data points par decade, i.e., $\Delta\ell/\ell = 0.787$.
   The spectra are computed using halofit formula \citep{2003MNRAS.341.1311S} with revised parameters by 
   \citet{2012ApJ...761..152T}.
   \label{fig:r_photoz}}
\end{figure}

In order to quantify this overlap, we compute the cross power spectrum between the two profiles.
Since the cross spectrum is expected to be zero in the ideal situation of $\sigma_0=0$,
it provides us a measure of the accuracy of nulling.
It is convenient to introduce the cross correlation coefficient between the two profiles:
\begin{eqnarray}
r_\ell^{12} = \frac{C_\ell^{12}}{\sqrt{C_\ell^{11}C_\ell^{22}}}.
\end{eqnarray}
Although $C_\ell^{12}$ itself is dependent on the normalization of the weight functions $p_1(z)$ and $p_2(z)$, 
which can  be chosen arbitrarily, the coefficient $r_\ell^{12}$ is not and it quantifies the relative amplitude of the
cross power spectrum to the auto power spectra.
We show this coefficient in Fig.~\ref{fig:r_photoz} when we adopt the same values of $\sigma_0$ as in 
Fig.~\ref{fig:kernel_photoz}.
In computing $C_\ell^{ij}$, we adopt the revised halofit.

When $\sigma_0$ is $0.03$, which is the typical target accuracy in future projects,
the coefficient is $\simlt10^{-3}$. It means that most of the lensing signal from the two profiles lies in the auto power
spectra with this value of $\sigma_0$. The coefficient can be as large as $10^{-2}$ to $10^{-1}$ when the dispersion of 
the photometric redshift is $\sim10\%$ depending on the multipole $\ell$ only weakly. 

For comparison,  the shaded region in Fig.~\ref{fig:r_photoz} locates the level of
the expected statistical error on this coefficient, $\Delta C_\ell^{12}/\sqrt{C_\ell^{11}C_\ell^{22}}$, for
a survey with a source number density of $n_\mathrm{tot} = 40\,\mathrm{arcmin}^{-2}$ 
in a survey area of $20,000\,\mathrm{deg}^2$.
We estimate this error from
\begin{eqnarray}
[\Delta C_\ell^{12}]^2 &=& \frac{1}{N^\mathrm{mode}_\ell}\left[(C_\ell^{11}+C^{11}_\mathrm{shape})(C_\ell^{22}+C^{22}_\mathrm{shape})\right.\nonumber\\
&&\hspace{2cm}+\left.(C_\ell^{12}+C^{12}_\mathrm{shape})^2\right],
\label{eq:error12}
\end{eqnarray}
where $N^\mathrm{mode}_\ell$ denotes the number of modes,
\begin{eqnarray}
N^\mathrm{mode}_\ell = 2 f_\mathrm{sky} (\ell+1) \Delta\ell,
\label{eq:Nmode}
\end{eqnarray}
that depends on the fraction of the observed sky $f_\mathrm{sky}$ and the size of the $\ell$-bin $\Delta\ell$.
In the above, $C^{ij}_\mathrm{shape}$ denotes the shape noise power spectrum:
\begin{eqnarray}
C^{ij}_\mathrm{shape} = \frac{\sigma^2_\gamma}{n_\mathrm{tot}}\int_0^{z_\infty} p_i(z_p)p_j(z_p)
n_p(z_p)\mathrm{d}z_p,
\label{eq:Cshape}
\end{eqnarray}
where $\sigma_\gamma$ is the dispersion of the individual galaxy shape and 
$n(z_p)$ is the normalized distribution function of the photometric redshift, that can be computed from $n(z)$
for a given $\sigma_0$.

We adopt the value $\sigma_\gamma = 0.22$ and the redshift distribution of the source in Eq.~(\ref{eq:nz}) in this calculation.
Note that although the derivation of the formula (\ref{eq:error12})  is based on the Gaussianity of the convergence field 
\citep{1994ApJ...426...23F}, it is still exact even when there is non-Gaussianity as long as nulling is exact.
This is because the cross trispectrum of the two profiles disappears thanks to the nulling property of the one
restricted in $1<z<2$.
In evaluating Eq.~(\ref{eq:error12}), we consider only the first term.
The shape noise power spectrum $C_\mathrm{shape}^{12}$ equals to zero, because no source galaxy
is included in both the profiles regardless of the value of $\sigma_0$.
In addition, we consider the situations where nulling is approximately implemented (i.e., $C^{12}_\ell\simeq0$).
Thus, the first term in Eq.~(\ref{eq:error12}) is dominant over the second term, and the latter can safely be neglected.

The resultant statistical error on the cross correlation coefficient, $r_\ell^{12}$, plotted in Fig.~\ref{fig:r_photoz}
is of the same order of magnitude as the cross correlation signal when $\sigma_0\simeq0.1$.
If the dispersion of the photometric redshift is unexpectedly large and is around $10\%$, 
our technique might be useful to detect it.
For a target redshift accuracy of $\sigma_0=0.03$, the signal is much smaller than the error level.
Provided the errors on the photometric redshift are properly controlled, one should then be able to safely
implement nulling within the statistical errors of future projects.

\begin{figure}
   \centering
 \includegraphics[width=7cm]{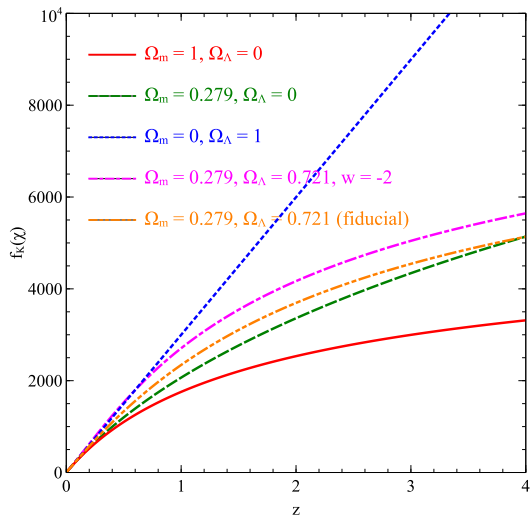}
   \caption{Comoving Angular distance in unit of $h^{-1}\mathrm{Mpc}$
   as a function of redshift for five models we consider here.
   Matter density as well as that of cosmological constant are indicated in the labels.
   The model plotted in dot-dashed line considers a more general dark energy model with the equation-of-state
   parameter equals to $-2$, and the rest of the models are $\Lambda$CDM.
   Our fiducial model is shown in dot-dot-dashed line.
   \label{fig:dz}}
\end{figure}

\begin{figure}
   \centering
 \includegraphics[width=8.5cm]{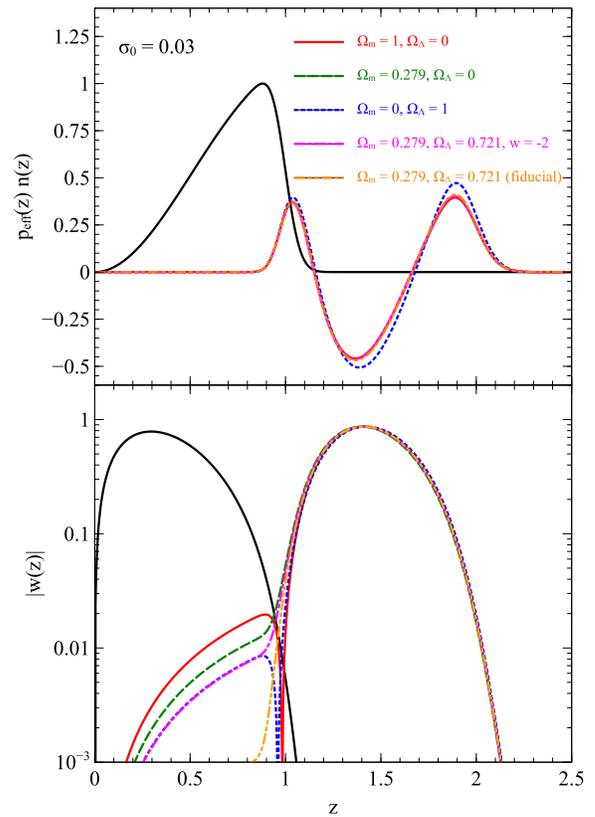}
   \caption{Effect of a wrong cosmological assumption on the nulling. Two adjacent profiles (thin: $0<z<1$, 
   thick: $1<z<2$) are plotted for different cosmological models as shown in the legend. Top: the weighted 
   source distribution. Bottom: the resultant lensing profile.
   Note that $p_\mathrm{eft}(z)n(z)$ ($w(z)$) has a dimension of 1/length (length), 
   and their normalization can be taken arbitrary.
   \label{fig:kernel_cosmo}}
\end{figure}

We then turn to the discussion on the error of nulling induced by a wrong assumption in the geometry of the universe.
We test the accuracy of nulling with a choice of five different cosmological models 
when the correct cosmology is a flat $\Lambda$CDM model with $\Omega_\mathrm{m}=0.279$.
The five models we consider are rather extreme cases; they are listed on Fig.~\ref{fig:dz} which shows their resulting angular diameter 
distances as a function of redshift.
We then test the accuracy of the nulling method by constructing the a priori nulling profiles assuming the various
cosmological models and then examining the resulting profile in the actual cosmology.
As in the previous paragraph, we use  two adjacent profiles.
The weighted number density of the source galaxies as well as the absolute value of the resultant lensing profile are plotted 
in Fig.~\ref{fig:kernel_cosmo}. We adopt $\sigma_0=0.03$ for the photometric redshift dispersion in this plot.
Non-negligible leakage of lensing profile can be observed at $z<1$ except for the choice of the fiducial cosmology.

\begin{figure}
   \centering
 \includegraphics[width=7cm]{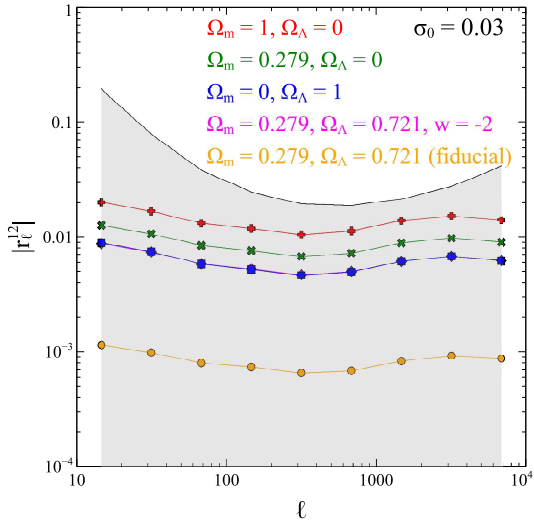}
   \caption{Cross correlation coefficient, $r_\ell^{12}$, when a wrong cosmological model is adopted in
   computing the weight function. The expected error level on $r_\ell^{12}$ shown in shade are the same as in
   Fig.~\ref{fig:r_photoz}.
   The spectra are computed using halofit formula \citep{2003MNRAS.341.1311S} with revised parameters by 
   \citet{2012ApJ...761..152T} as in Fig.~\ref{fig:r_photoz}.}
   \label{fig:r_cosmo}
\end{figure}

We finally show in Fig.~\ref{fig:r_cosmo} the cross-correlation coefficient of the two profiles obtained with the five cosmological
models. The standard CDM model gives the largest signal among the five, and is close to the noise level
given by Eq.~(\ref{eq:error12}) with the same survey design as before.
This gives a rough estimate of the upper limit of the failed nulling signal. It suggests that we can safely implement nulling 
with more realistic cosmological assumptions. Note however that 
although we cannot detect a statistically meaningful signal when we focus on each of the $\ell$-bins, we might be
able to detect it by combining several bins and using multiple nulling profiles, and eventually falsify the assumed 
cosmological model from such a diagnosis alone.
This feature can be used as a unique test of cosmology, which provides us purely geometrical constraints.
A more thorough discussion of the constraining power of the cosmological models through the measurement of
this failed null signal will be given elsewhere together with the optimal design of a set of profiles to cover the 
whole range of redshift.

\subsection{Applicable range of perturbation theories}

With a successful construction of the nulling lensing profile 
in a realistic case, we discuss here the impact of this technique on the practical 
application of perturbation theory, 
just reconsidering the results of Sec.~\ref{Sec:PTresults}.

For a continuous source distribution, using Eq.~(\ref{eq:nchishape1})
we can construct any nulling profile with an arbitrary 
redshift interval. With a sufficiently large number of source galaxies, 
the redshift interval of the nulling profile can be made arbitrarily 
narrow so that the lensing kernel $w(\chi)$ is approximately described 
by $w(\chi)\to\delta_{\rm D}(\chi-\chi_s)$, where $\chi_s$ is the radial 
distance to the source galaxies at redshift $z_s$. 
In this case, the multipole of the lensing power spectrum $C_\ell$ is 
directly related to the wavenumber of the 
three-dimensional power spectrum $P(k)$ at $z_s$ through
\begin{eqnarray}
\ell\simeq k \,\,f_{\rm K}(\chi_s).
\label{eq:k_ell}
\end{eqnarray}
Thus, the accessible range of perturbation theory in $k$-space 
is simply mapped into the one in multipole. This is to be contrasted with the 
case without nulling technique. Even using source galaxies localized within 
an infinitesimally narrow redshift interval
the contribution from the small-scale nonlinearity can affect 
the lensing power spectrum through the projection effect as shown in 
Fig.~\ref{kernelklprof4}, 
shrinking the applicable range of perturbation theory. 

\begin{figure}
   \centering
 \includegraphics[width=8.5cm]{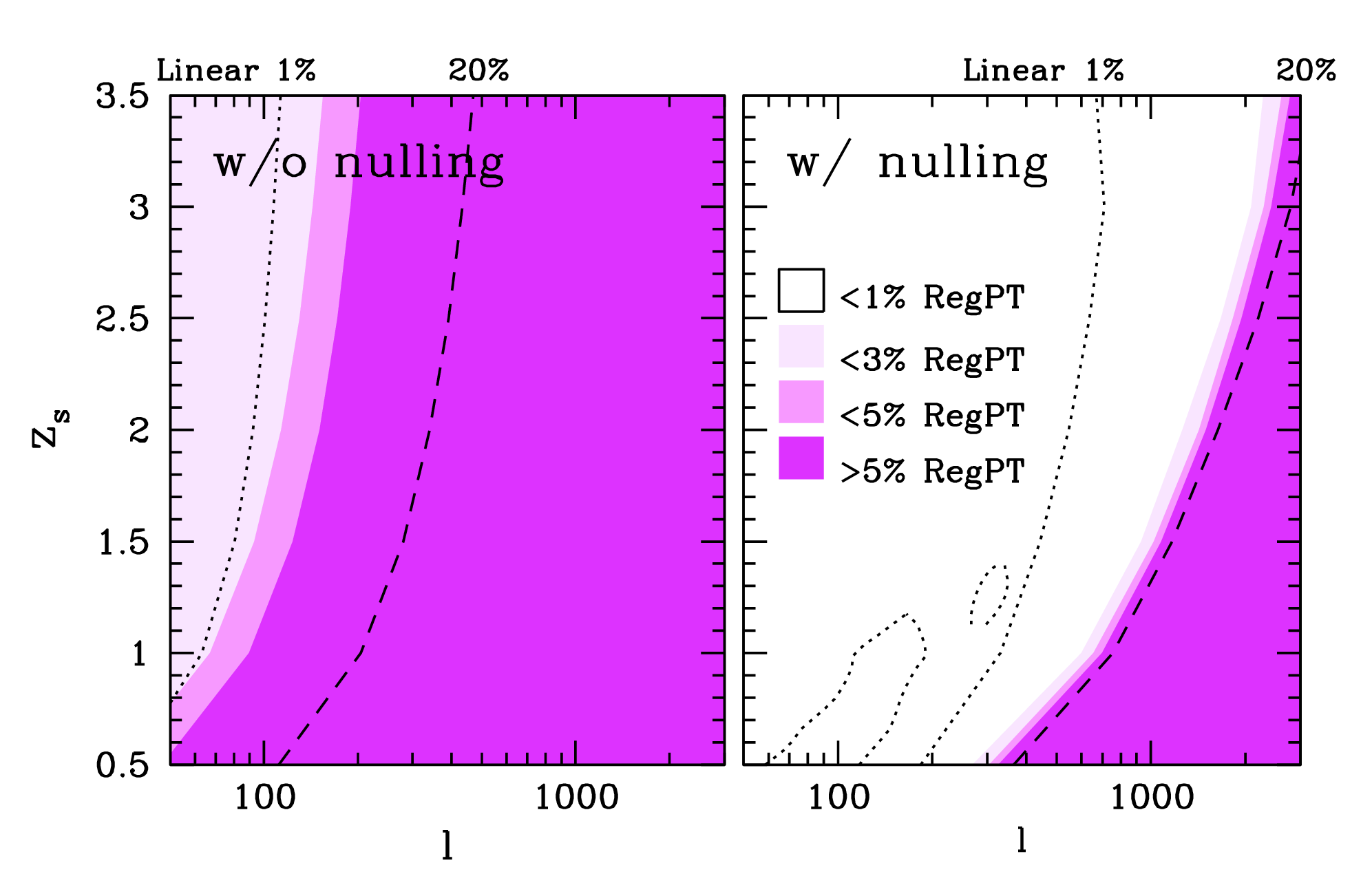} 
   \caption{
   Impact of small-scale nonlinearity on the lensing power spectrum 
with (right) and without (left) nulling technique. Shaded region indicate
the size of fractional difference between RegPT predictions and 
nonlinear power spectrum at different source redshift, $z_s$, plotted as 
a function of multipole $\ell_{\rm lim}$; 
$1\%$, $3\%$, and $5\%$ from lighter to darker. 
Here, we assume the best-fit Planck cosmology, and the reference nonlinear power 
spectrum is computed with an updated version of the 
cosmic emulator code that provides interpolated 
power spectra from high-resolution $N$-body simulations. We convert the results in wavenumber $k$ to those
in multipole $\ell$ using Eq.~(\ref{eq:k_ell}).
The dotted lines
represent the locations of the $1\%$ accuracy domain of the linear theory predictions and 
the dashed line the $20 \%$ accuracy domain. In case of nulling it approximately gives the extent of
the RegPT validity domain.
}
   \label{elllimit}
\end{figure}

Fig.~\ref{elllimit} summarizes the impact of small-scale nonlinearity on the 
lensing power spectrum 
with (right) and without (left) nulling technique. The shaded colors indicate
the fractional difference between nonlinear power spectrum and the perturbation 
theory prediction at different source redshift, $z_s$, as a function of 
multipole $\ell_{\rm lim}$. Here, we assume the best-fit Planck cosmology
\citep{2013arXiv1303.5076P}
and 
the reference nonlinear power spectrum is computed with an updated version of the 
cosmic emulator code that provides interpolated 
power spectra from high-resolution $N$-body simulations 
\citep{2013arXiv1304.7849H}.
For perturbation theory prediction, we adopt the 
RegPT at two-loop order as a representative resummed PT technique
\citep{2012PhRvD..86j3528T}.
We determine the wavenumber ranges of the RegPT as well as the linear theory 
by confronting predictions with those of the cosmic emulator.
Taking advantage of the continuous source distribution, we consider 
the idealistic situation as discussed above, 
and pick up the source galaxies at arbitrary $z_s$ with 
infinitesimally thin redshift interval. In this case, with nulling technique, 
a simple relation with Eq.~(\ref{eq:k_ell}) may be applied 
to estimate the fractional discrepancy (right).

Fig.~\ref{elllimit} clearly shows that the nulling technique is very 
powerful to mitigate 
the impact of small-scale nonlinearity. Without the nulling technique, 
the small-scale nonlinearities are not controlled within PT calculations, making
the accessible range of RegPT results even narrower than 
that of the linear theory predictions (dotted and dashed lines).
However, the situation is dramatically changed if we consider the 
nulling technique. The reliable range of 
RegPT predictions becomes much wider as shown 
by the location of the shaded areas in the right panel.
The accessible range of RegPT 
prediction at $1\%$ precision now extends 
over $\ell=1300-2100$ at higher source redshift $z_s=2-3$.   
We found that this is roughly comparable to the scale where linear theory prediction 
produces the $20\%$ error (dashed line). Note that with the nulling technique, 
even the linear theory can give a reliable prediction 
at $1\%$ precision (dotted line) up to $\ell=550-700$ at 
redshift $z_s=2-3$, (note that at lower redshift $z_s\lesssim1.5$, as
the nonlinear growth of structure deforms the BAO structure, the 
$1\%$ boundary line, depicted as 
dotted lines, is made convoluted).
This is a dramatic improvement. 
Of course, in practice, shot-noise 
contribution can be large due to the finite number of source galaxies, 
and the lens distribution will have a finite width. 
Nevertheless, this simple demonstration gives us a  
useful and general guideline to the extent with which we can apply 
perturbation theory to  weak lensing experiments.


\section{Conclusions}

We have presented a nulling construction that allows to reorganize tomographic information in such a way that 
different contributions to the shear maps are sorted out, geometry, systematics and regimes of dynamical evolution, 
a property that standard tomographic constructions does not exhibit.

After such a transformation, the correlation matrix between different maps is indeed band diagonal for all $\ell$, 
and most cross-correlations are nulled. This information can be exploited to all scales to constrain basic cosmological parameters, those
related to the geometrical parameters. The idea we have developed here is based on the possibility of having lens 
distributions confined to a finite, and possibly narrow, range of redshift. Following what we have developed in the paper,
such a scheme offers two advantages,
	\begin{itemize} 
	\item the nulling is valid irrespectively of the regime - linear or nonlinear - 
and this is a key property. That means that one can use the nulling information with its full power even in regimes 
where exact analytic prediction are difficult;
	\item because one can select the redshift range of the lenses, for each chosen bin, angular scales are more closely related 
	to physical scales making it easier to make analytical predictions. In particular because linear and nonlinear scales are not 
	mixed up one can obtain controlled predictions to higher $\ell$ for specific source choices.
	\end{itemize}
Note that in terms of amount of information there is no less and no more than with standard tomography. However the information 
is somehow sorted out in terms of theoretical,  astrophysical and instrumental systematics. More specifically we can then put forward 
the part of the data that are free of theoretical uncertainties (i.e. for which one can compute exactly the statistical properties). 
Observed correlation when nulled signal is expected  could then be used as a way to track down systematics errors such 
instrumental systematics (through seeing, pixellisation, masking) or astrophysical systematics through intrinsic alignment effects.
Note incidentally that the fact that we have the full $\ell$ dependence of the cross-spectra  should help sorting out those effects.

Furthermore the band diagonal elements themselves are better behaved in the sense that they are less sensitive to projection effects.
As a result the mapping between $\ell$ and $k$ is much more precise (as illustrated on Fig. \ref{kernelklprof4}) making possible to
associate, for each map, more closely angular scales to physical scales.  And least but not last, it allows to make predictions
from perturbation theory calculations to smaller angular scales, and all the more smaller that maps correspond to
more distant lenses. The accuracy of such predictions are shown in Sect. \ref{Sec:PTresults}. They show that analytical calculations can
account of cosmic shear spectra up to $\ell$ about 1000 when lenses are at about redshift unity.

\section*{Acknowledgments}
We appreciate Masanori Sato for kindly providing us with the convergence maps constructed by ray-tracing simulations. 
This work is partially supported by grant ANR-12-BS05-0002 of the French Agence Nationale de la Recherche
and by Grant-in-Aid for Scientific Research from the JSPS  (No.~24540257 for AT).
TN is supported by Japan Society for the Promotion of Science (JSPS) Postdoctoral Fellowships
for Research Abroad. 
FB also thanks the YITP of the university of Kyoto for hospitality during the completion of this work. 
\bibliographystyle{mn2e}
\bibliography{Nulling}

\begin{thebibliography}{}
 \providecommand{\href}[2]{#2}
  \providecommand{\doi}[1]{\href{http://dx.doi.org/#1}{doi:#1}}
  \providecommand{\eprint}[1]{\href{http://arxiv.org/abs/#1}{arXiv:#1}}

\bibitem[\protect\citeauthoryear{{Bacon}, {Refregier} \& {Ellis}}{{Bacon}
  et~al.}{2000}]{2000MNRAS.318..625B}
{Bacon} D.~J.,  {Refregier} A.~R.,    {Ellis} R.~S.,  2000, \mnras, 318, 625,
  \eprint{arXiv:astro-ph/0003008}

\bibitem[\protect\citeauthoryear{{Crocce} \& {Scoccimarro}}{{Crocce} \&
  {Scoccimarro}}{2006}]{2006PhRvD..73f3519C}
{Crocce} M.,  {Scoccimarro} R.,  2006, \prd, 73, 063519,
  \eprint{arXiv:astro-ph/0509418}

\bibitem[\protect\citeauthoryear{{Crocce}, {Scoccimarro} \&
  {Bernardeau}}{{Crocce} et~al.}{2012}]{2012MNRAS.427.2537C}
{Crocce} M.,  {Scoccimarro} R.,    {Bernardeau} F.,  2012, \mnras, 427, 2537,
  \eprint{1207.1465}

\bibitem[\protect\citeauthoryear{{Feldman}, {Kaiser} \& {Peacock}}{{Feldman}
  et~al.}{1994}]{1994ApJ...426...23F}
{Feldman} H.~A.,  {Kaiser} N.,    {Peacock} J.~A.,  1994, \apj, 426, 23,
  \eprint{astro-ph/9304022}

\bibitem[\protect\citeauthoryear{{Fu}, {Semboloni}, {Hoekstra}, {Kilbinger} \&
  {et al.}}{{Fu} et~al.}{2008}]{2008A&A...479....9F}
{Fu} L.,  {Semboloni} E.,  {Hoekstra} H.,  {Kilbinger} M.,    {et al.} 2008,
  \aap, 479, 9, \eprint{0712.0884}

\bibitem[\protect\citeauthoryear{{Heavens}}{{Heavens}}{2003}]{2003MNRAS.343.13%
27H}
{Heavens} A.,  2003, \mnras, 343, 1327, \eprint{astro-ph/0304151}

\bibitem[\protect\citeauthoryear{{Heavens}, {Kitching} \& {Taylor}}{{Heavens}
  et~al.}{2006}]{2006MNRAS.373..105H}
{Heavens} A.~F.,  {Kitching} T.~D.,    {Taylor} A.~N.,  2006, \mnras, 373, 105,
  \eprint{astro-ph/0606568}

\bibitem[\protect\citeauthoryear{{Heitmann}, {Lawrence}, {Kwan}, {Habib} \&
  {Higdon}}{{Heitmann} et~al.}{2013}]{2013arXiv1304.7849H}
{Heitmann} K.,  {Lawrence} E.,  {Kwan} J.,  {Habib} S.,    {Higdon} D.,  2013,
  ArXiv e-prints, \eprint{1304.7849}

\bibitem[\protect\citeauthoryear{{Heymans} et~al.,}{{Heymans}
  et~al.}{2012}]{2012MNRAS.427..146H}
{Heymans} C.  et~al., 2012, \mnras, 427, 146, \eprint{1210.0032}

\bibitem[\protect\citeauthoryear{{Huterer} \& {White}}{{Huterer} \&
  {White}}{2005}]{2005PhRvD..72d3002H}
{Huterer} D.,  {White} M.,  2005, \prd, 72, 043002, \eprint{astro-ph/0501451}

\bibitem[\protect\citeauthoryear{{Hu}}{{Hu}}{1999}]{1999ApJ...522L..21H}
{Hu} W.,  1999, \apjl, 522, L21, \eprint{arXiv:astro-ph/9904153}

\bibitem[\protect\citeauthoryear{{Joachimi} \& {Schneider}}{{Joachimi} \&
  {Schneider}}{2008}]{2008A&A...488..829J}
{Joachimi} B.,  {Schneider} P.,  2008, \aap, 488, 829, \eprint{0804.2292}

\bibitem[\protect\citeauthoryear{{Kitching} \& {Taylor}}{{Kitching} \&
  {Taylor}}{2011}]{2011MNRAS.416.1717K}
{Kitching} T.~D.,  {Taylor} A.~N.,  2011, \mnras, 416, 1717, \eprint{1012.3479}

\bibitem[\protect\citeauthoryear{{Kitching}, {Heavens} \& {Miller}}{{Kitching}
  et~al.}{2011}]{2011MNRAS.413.2923K}
{Kitching} T.~D.,  {Heavens} A.~F.,    {Miller} L.,  2011, \mnras, 413, 2923,
  \eprint{1007.2953}

\bibitem[\protect\citeauthoryear{{Laureijs} et~al.,}{{Laureijs}
  et~al.}{2011}]{2011arXiv1110.3193L}
{Laureijs} R.  et~al., 2011, ArXiv e-prints, \eprint{1110.3193}

\bibitem[\protect\citeauthoryear{{Mellier}}{{Mellier}}{1999}]{1999ARA&A..37..1%
27M}
{Mellier} Y.,  1999, \araa, 37, 127

\bibitem[\protect\citeauthoryear{{Pietroni}}{{Pietroni}}{2008}]{2008JCAP...10.%
.036P}
{Pietroni} M.,  2008, \jcap, 10, 36, \eprint{0806.0971}

\bibitem[\protect\citeauthoryear{{Planck Collaboration}}{{Planck
  Collaboration}}{2013}]{2013arXiv1303.5076P}
{Planck Collaboration} 2013, ArXiv e-prints, \eprint{1303.5076}

\bibitem[\protect\citeauthoryear{{Sato}, {Hamana}, {Takahashi}, {Takada},
  {Yoshida}, {Matsubara} \& {Sugiyama}}{{Sato}
  et~al.}{2009}]{2009ApJ...701..945S}
{Sato} M.,  {Hamana} T.,  {Takahashi} R.,  {Takada} M.,  {Yoshida} N.,
  {Matsubara} T.,    {Sugiyama} N.,  2009, \apj, 701, 945, \eprint{0906.2237}

\bibitem[\protect\citeauthoryear{{Smith} et~al.,}{{Smith}
  et~al.}{2003}]{2003MNRAS.341.1311S}
{Smith} R.~E.  et~al., 2003, \mnras, 341, 1311, \eprint{astro-ph/0207664}

\bibitem[\protect\citeauthoryear{{Takahashi}, {Sato}, {Nishimichi}, {Taruya} \&
  {Oguri}}{{Takahashi} et~al.}{2012}]{2012ApJ...761..152T}
{Takahashi} R.,  {Sato} M.,  {Nishimichi} T.,  {Taruya} A.,    {Oguri} M.,
  2012, \apj, 761, 152, \eprint{1208.2701}

\bibitem[\protect\citeauthoryear{{Taruya} \& {Hiramatsu}}{{Taruya} \&
  {Hiramatsu}}{2008}]{2008ApJ...674..617T}
{Taruya} A.,  {Hiramatsu} T.,  2008, \apj, 674, 617, \eprint{0708.1367}

\bibitem[\protect\citeauthoryear{{Taruya}, {Bernardeau}, {Nishimichi} \&
  {Codis}}{{Taruya} et~al.}{2012}]{2012PhRvD..86j3528T}
{Taruya} A.,  {Bernardeau} F.,  {Nishimichi} T.,    {Codis} S.,  2012, \prd,
  86, 103528, \eprint{1208.1191}

\bibitem[\protect\citeauthoryear{{Valageas}, {Sato} \& {Nishimichi}}{{Valageas}
  et~al.}{2012a}]{2012A&A...541A.161V}
{Valageas} P.,  {Sato} M.,    {Nishimichi} T.,  2012a, \aap, 541, A161,
  \eprint{1111.7156}

\bibitem[\protect\citeauthoryear{{Valageas}, {Sato} \& {Nishimichi}}{{Valageas}
  et~al.}{2012b}]{2012A&A...541A.162V}
{Valageas} P.,  {Sato} M.,    {Nishimichi} T.,  2012b, \aap, 541, A162,
  \eprint{1112.1495}

\bibitem[\protect\citeauthoryear{{Van Waerbeke} et~al.,}{{Van Waerbeke}
  et~al.}{2000}]{2000A&A...358...30V}
{Van Waerbeke} L.  et~al., 2000, \aap, 358, 30, \eprint{arXiv:astro-ph/0002500}

\bibitem[\protect\citeauthoryear{{Wittman}, {Tyson}, {Kirkman}, {Dell'Antonio}
  \& {Bernstein}}{{Wittman} et~al.}{2000}]{2000Natur.405..143W}
{Wittman} D.~M.,  {Tyson} J.~A.,  {Kirkman} D.,  {Dell'Antonio} I.,
  {Bernstein} G.,  2000, \nat, 405, 143, \eprint{arXiv:astro-ph/0003014}

\end{thebibliography}

\appendix

\section{Different prescriptions for the signal-to-noise ratio to construct continuous profiles}
\label{app:prescription}
In the main text, we adopt Eq.~(\ref{eq:nchishape1}) to obtain a smooth profile that maximizes the signal~(\ref{eq:S1})
with respect to the noise~(\ref{eq:N}). In this Appendix, we give two alternative prescriptions for the signal to noise, and
show that the resultant profiles are not sensitive to the detail of the prescription.

\begin{figure}
   \centering
 \includegraphics[width=8cm]{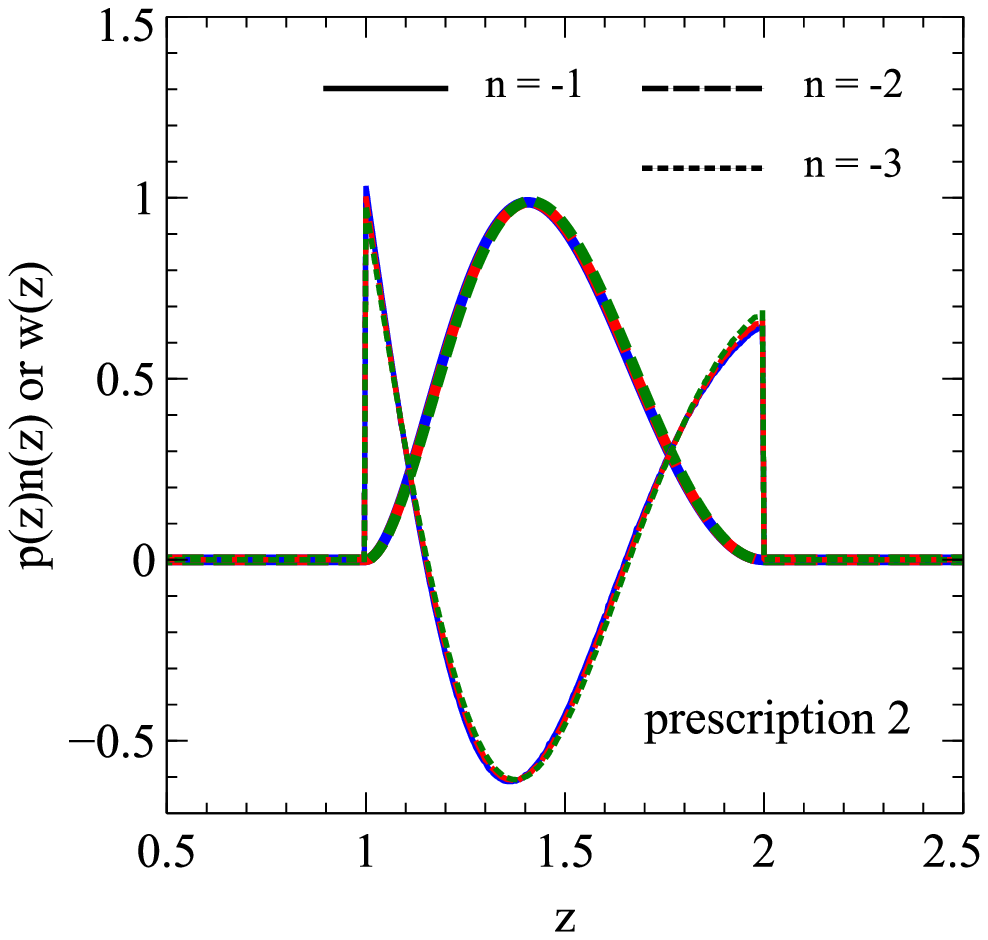}
   \caption{Nulling profiles obtained with the maximization of $(\mS_2/\mN)$.
   Note that $p(z)n(z)$ ($w(z)$) has a dimension of 1/length (length), 
   and their normalization can be taken arbitrary.
   \label{fig:prescription2}}
\end{figure}

An alternative, and better justified approach, is to define the signal based on the significance of the power 
spectrum~(\ref{eq:power}). Assuming that the local convergence can be estimated using the linear theory
where the density power spectrum evolves linearly and scales like $a^{2}(\chi)$ and the wavenumber dependence of
the three-dimensional power spectrum is effectively given by a power law with index $n$, $P(k)\propto k^n$, 
we define the signal by
\begin{eqnarray}
\mS_{2}^2&=&\int_{0}^{\chi_{\infty}}\dd\chi
\,\fK(\chi)^{-(n+2)}w^2(\chi),
\end{eqnarray}
where the lensing profile $w(\chi)$ is given by Eq.~(\ref{eq:wkernel}).
The profiles obtained by maximization of $\mS_2 / \mN$ are plotted in Fig.~\ref{fig:prescription2} for
three values of the effective spectral index, $n=-1$, $-2$ and $-3$. The dependence of the profile on the parameter 
$n$ is rather weak.

Although it requires a model for the nonlinear power spectrum and its covariance property,
we might introduce another definition of signal to noise, which is more related to the accessible information 
content from a power spectrum analysis. We define
\begin{eqnarray}
\left(\frac{\mS_3}{\mN}\right)^2 = \sum_{\ell,\ell'}C_\ell\left(\mathrm{Cov}_{\ell\ell'}\right)^{-1} C_{\ell'},
\end{eqnarray}
where $\mathrm{Cov}_{\ell,\ell'}$ denotes the covariance between $C_\ell$ and $C_{\ell'}$. Assuming Gaussianity of
the field $\kappa$ and for given survey parameters this reduces to
\begin{eqnarray}
\left(\frac{\mS_3}{\mN}\right)^2 = f_\mathrm{sky}\sum_{\ell<\ell_\mathrm{max}}\frac{2\ell+1}{2}
\left[1+\frac{C_\mathrm{shape}}{C_\ell}\right]^{-2},
\end{eqnarray}
where we denote by $\ell_\mathrm{max}$ the maximum multipole taken into the summation, and the shape noise is given by
\begin{eqnarray}
C_\mathrm{shape} = \frac{\sigma_\gamma^2}{n_\mathrm{tot}}\int_{\chi_{1}}^{\chi_{2}}\dd\chi_{s}\,p^{2}(\chi_{s})n(\chi_{s})
\propto \mN^{2},
\end{eqnarray}
anologously to Eq.~(\ref{eq:Cshape}).

\begin{figure}
   \centering
 \includegraphics[width=8cm]{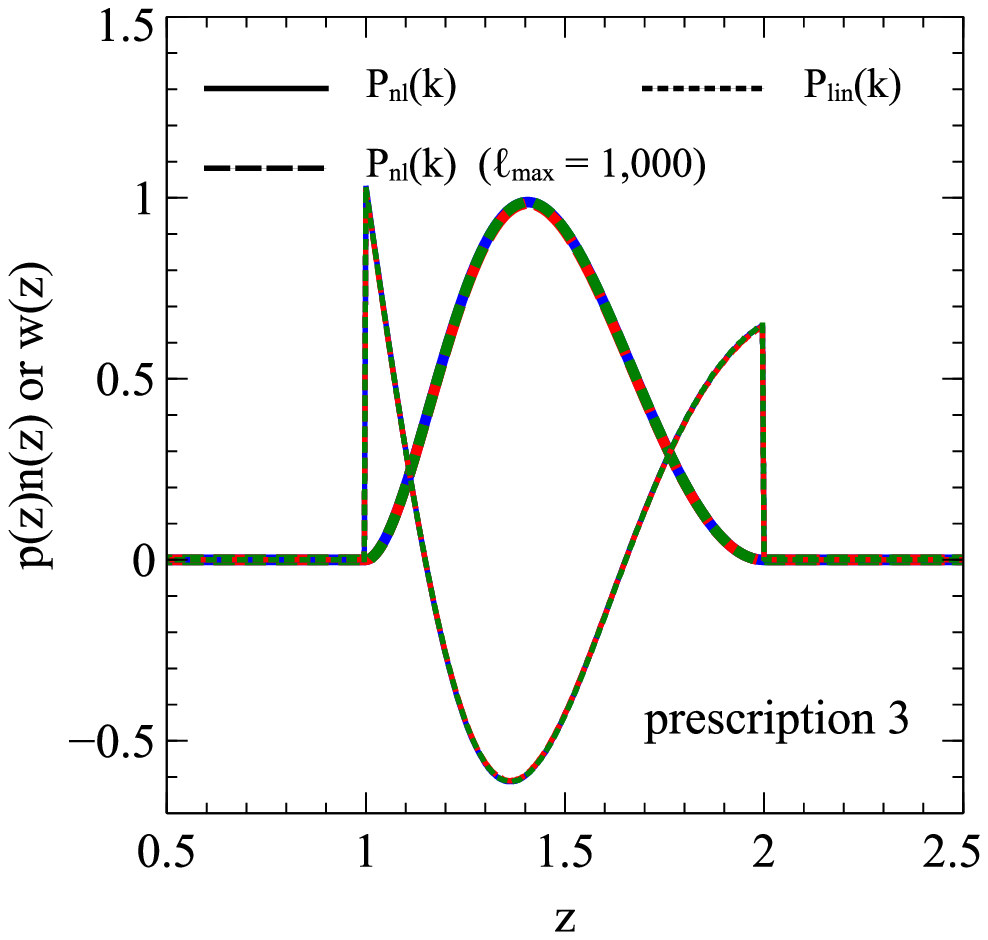}
   \caption{Nulling profiles obtained with the maximization of $(\mS_3/\mN)$.
   Note that $p(z)n(z)$ ($w(z)$) has a dimension of 1/length (length), 
   and their normalization can be taken arbitrary.
   \label{fig:prescription3}}
\end{figure}

The resulting shape of $p(\chi_{s})$ and the lens distribution functions are shown in Fig. \ref{fig:prescription3}.
We employ the fitting formula of the nonlinear power spectrum given by \cite{2012ApJ...761..152T} for solid and
dashed line, while the linear power spectrum is used for the dashed line. Also, we adopt $\ell_\mathrm{max} = 10,000$
except for the dashed line, which adopts $\ell_\mathrm{max}=1,000$.
Again, we can see that the dependence of the profile on the detail of the model is rather weak.

\begin{figure}
   \centering
 \includegraphics[width=8cm]{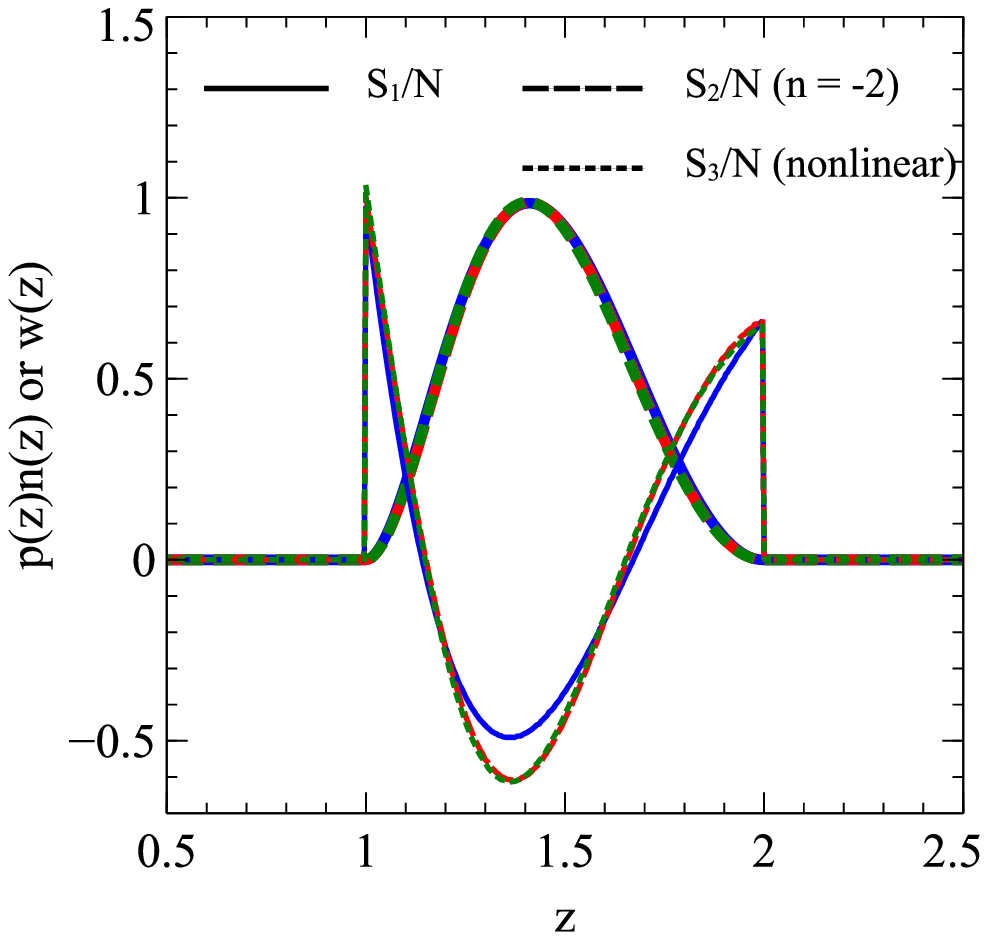}
   \caption{Comparison of the nulling profiles with different prescriptions for the signal to noise maximization. 
   Note that $p(z)n(z)$ ($w(z)$) has a dimension of 1/length (length), 
   and their normalization can be taken arbitrary.
   \label{fig:prescription_comp}}
\end{figure}

We finish this Appendix with a comparison of the profiles obtained with the three different prescriptions of signal to noise
as shown on 
Fig.~\ref{fig:prescription_comp}.
To compute $\mS_2$ ($\mS_3$), we adopt $n=-2$ (the nonlinear matter power spectrum up to $\ell_\mathrm{max} = 10,000$).
Notice the similarity of the profiles obtained with the maximization of $\mS_2/\mN$ and $\mS_3/\mN$.
Although the prescription adopted in the main text exhibits a slightly shallower dip of the weight function at $z\sim1.4$,
the resulting profile, $w$, are almost indistinguishable.

\section{Effect of simulation window on the power spectrum measurement}
\label{app:window}
In this paper, we assess the validity range of the perturbation theories by confronting it with numerical simulations.
Although the analytical estimate of the nonlinear power spectrum is expected to be more accurate at larger scale
(i.e., at smaller $\ell$), one may notice a slight, but statistically significant discrepancy with the numerical results at $\ell\simlt200$ 
(see figure~\ref{SatoCl_Nulling}). This feature has also been reported in previous studies using the same numerical
simulations \citep{2012A&A...541A.161V,2012A&A...541A.162V}. 
Since the accuracy of the models are ultimately  justified by their consistency with simulations, it is important to
fully understand the reason of this discrepancy.

We find out that this is likely due to the effect of finite area of ray-tracing simulations.
Although the simulations we use have $25,000\,\mathrm{deg}^2$ in total of $1,000$ realizations,
each simulated map covers only an area of $5\,\mathrm{deg}\times5\,\mathrm{deg}$.
Since one cannot mitigate the window effect by increasing the number of realizations to be averaged over,
the final estimate of the power spectrum shows a slight underestimate of power at large scales comparable to the size
of the each simulated convergence map. 

We may write the convergence field obtained in simulations as
\begin{eqnarray}
\kappa_\mathrm{w}(\mathbf{\theta}) = \int \mathrm{d}^2\theta' W(\mathbf{\theta}-\mathbf{\theta}')\kappa(\mathbf{\theta'}),
\end{eqnarray}
where the window function $W(\mathbf{\theta})$ is unity inside the simulation area while it is zero outside.
Then the power spectrum of the windowed field, $\kappa_\mathrm{w}$, can be written as
\begin{eqnarray}
C^\mathrm{w}_\ell = \left|\widetilde{W}(\ell)\right|^2C_\ell,
\end{eqnarray}
where $\widetilde{W}$ denotes the Fourier transform of the window function $W$.

We compute the analytical power spectrum taking into account this convolution with the following procedure:
we first prepare a square area of $10\,\mathrm{deg} \times 10\,\mathrm{deg}$ with periodic boundary, and 
generate a Gaussian random field on $256\times256$ grid points that has the power spectrum $C_\ell$
computed with RegPT up to the 2-loop level.
We then clip a $5\,\mathrm{deg}\times5\,\mathrm{deg}$ region out of $10\,\mathrm{deg} \times 10\,\mathrm{deg}$,
and measure the power spectrum for the clipped region.
We repeat this procedure for $10,000$ times and take average of the power spectra over realizations to obtain
an estimate of $C^\mathrm{w}_\ell$.
We have checked that the result is stable against the area of the map in which we generate a Gaussian random field
or the number of grid points.

\begin{figure}
   \centering
 \includegraphics[width=8cm]{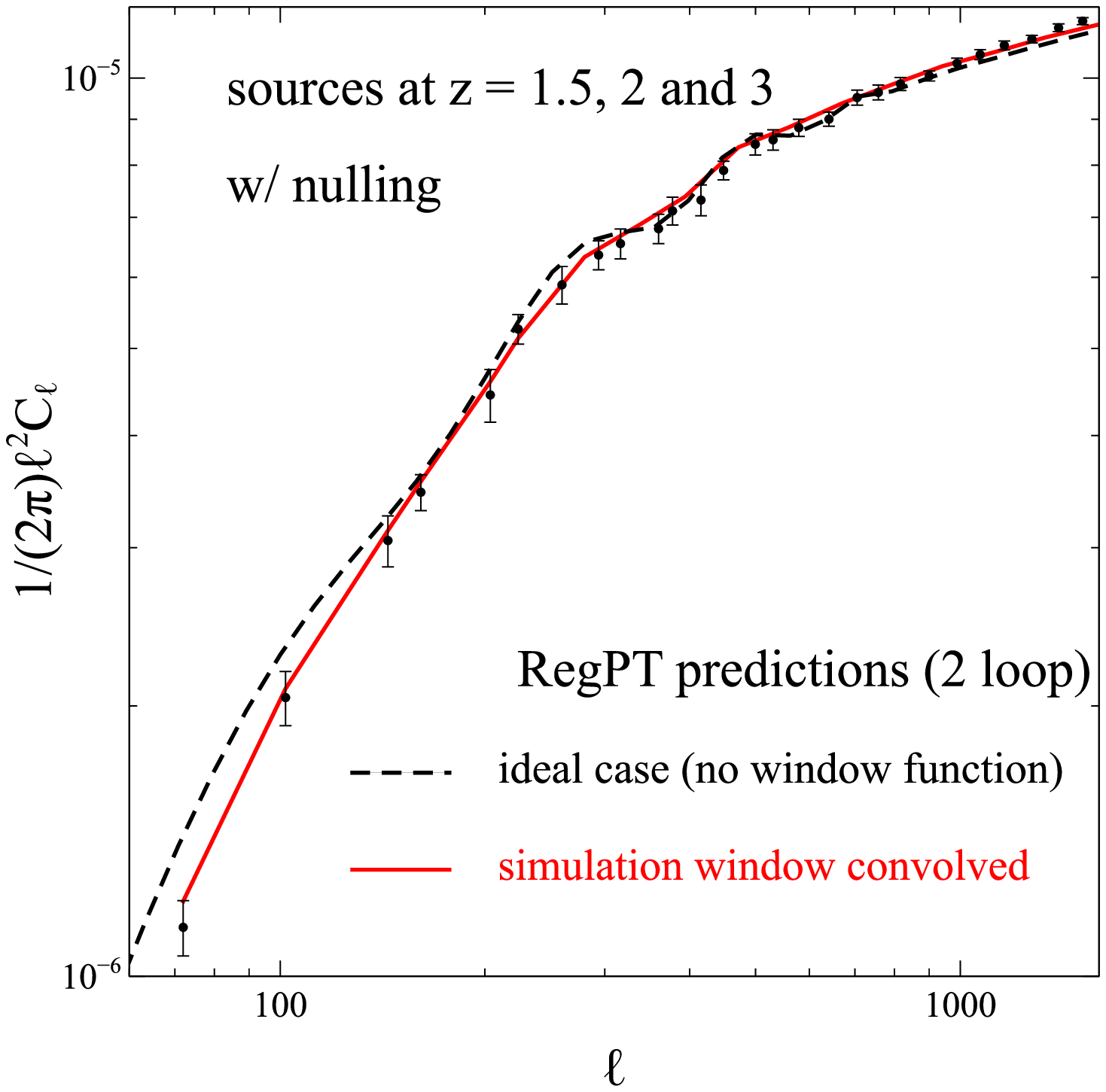}
   \caption{Comparison of the power spectrum with an without convolution of the window function.}
   \label{fig:window}
\end{figure}

The resultant analytical estimate is compared with simulations in Fig.~\ref{fig:window}.
We here use a nulling profile constructed from the three source planes at higher redshifts  
by \citet{2009ApJ...701..945S} in order to focus on linear to weakly nonlinear regime (the exact redshifts
of these source planes are $1.519$, $1.998$ and $3.057$).
We plot by solid (dashed) line the RegPT prediction with (without) a convolution of the window function.
The low-$\ell$ modes at $\ell\simlt200$ measured from simulations (symbols with error bars) are nicely explained 
by the solid line while the dashed line shows a poorer fit.
Another notable change induced by the convolution is the smoothed pattern of baryon acoustic oscillations seen at 
$300\simlt\ell\simlt700$. The simulation data again shows a good agreement with the solid line within the statistical error.

\end{document}